%% file: ms.tex
\documentclass[iop]{emulateapj}

\usepackage{amsmath}    

\usepackage{hyperref}
\usepackage{natbib}
\usepackage{color}
\usepackage{graphicx}
\usepackage{epstopdf}
\usepackage{graphicx}
\received{}
\accepted{}

\slugcomment{to appear in ApJ}
\shorttitle{}
\shortauthors{Wang et al.}

\begin{document}
\title{Time-Resolved High Spectral Resolution Observation of 2MASSW J0746425+200032AB}

\author{
Ji Wang\altaffilmark{1},
Lisa Prato \altaffilmark{2}, and
Dimitri Mawet\altaffilmark{1}
} 
\email{ji.wang@caltech.edu}
\altaffiltext{1}{Department of Astronomy, California Institute of Technology, MC 249-17, 1200 E. California Blv, Pasadena, CA 91106 USA}
\altaffiltext{2}{Lowell Observatory 1400 West Mars Hill Road Flagstaff, AZ 86001 }

\begin{abstract}

Many brown dwarfs exhibit photometric variability at levels from tenths to tens of percents. The photometric variability is related to magnetic activity or patchy cloud coverage, characteristic of brown dwarfs near the L-T transition. Time-resolved spectral monitoring of brown dwarfs provides diagnostics of cloud distribution and condensate properties. However, current time-resolved spectral studies of brown dwarfs are limited to low spectral resolution (R$\sim$100) with the exception of the study of Luhman 16 AB at resolution of 100,000 using the VLT$+$CRIRES. This work yielded the first map of brown dwarf surface inhomogeneity, highlighting the importance and unique contribution of high spectral resolution observations. Here, we report on the time-resolved high spectral resolution observations of a nearby brown dwarf binary, 2MASSW J0746425+200032AB. We find no coherent spectral variability that is modulated with rotation. Based on simulations we conclude that the coverage of a single spot on 2MASSW J0746425+200032AB is smaller than 1\% or 6.25\% if spot contrast is 50\% or 80\% of its surrounding flux, respectively. Future high spectral resolution observations aided by adaptive optics systems can put tighter constraints on the spectral variability of 2MASSW J0746425+200032AB and other nearby brown dwarfs.  

\end{abstract}


\section{Introduction}
\label{sec:intro}

As effective temperature ($\rm{T}_{\rm{eff}}$) drops, particles in the atmospheres of brown dwarfs (BDs) condense into dust or clouds. As a result, it is generally thought that BDs near the L-T transition ($1300 \rm{K} < \rm{T}_{\rm{eff}} < 2300 \rm{K}$) exhibit patchy clouds~\citep{Kirkpatrick1999, Burgasser2002}. As more exoplanets are discovered, the boundary between jovian planets and BDs becomes more blurred~\citep{Chen2016}. Similar to BDs, clouds on exoplanets influence their colors and spectra. Some of the observed spectra of exoplanets may only be explained by invoking clouds~\citep{Bonnefoy2015}. The issue is further complicated by disequilibrium chemistry and photochemistry for directly imaged exoplanets~\citep{Moses2016}. 

Because of surface non-uniformity, BDs and exoplanets with patchy clouds are characterized by photometric variability that arises as the result of their rotation. In order to understand cloud coverage, many photometric surveys were conducted which led to a large sample of BDs with photometric variability of tenths to tens of percents~\citep[][and references therein]{Metchev2015, Crossfield2014b}.  Spectroscopic monitoring of BDs reveals more about clouds on BDs than photometric surveys. However, most recent time-resolved spectroscopic studies are limited to low spectral resolution with a resolving power of $\sim$100. These observations imposed limits on particle sizes in BD atmospheres~\citep{Lew2016} and mapped surface spots and faculae coverage~\citep{Burgasser2014, Karalidi2016}.

Time-resolved, high-resolution, spectroscopic observations of BDs facilitate surface mapping via the Doppler imaging technique~\citep{Vogt1999}. By applying the technique to Luhman 16 AB, the nearest BDs~\citep{Luhman2013}, ~\citet{Crossfield2014} obtained the first high-fidelity surface map of Luhman 16 B. In principle, the Doppler imaging technique can also be applied to other BDs exhibiting large amplitude photometric variability and known rotational periods and axis orientation. Given sufficient signal to noise ratio (SNR), Doppler imaging of exoplanets will be feasible with next-generation instruments and telescopes. 

We report on the observations of 2MASSW J0746425+200032 AB (hereafter J0746 AB) with Keck II and NIRSPEC. This is the second example of a time-resolved, high-resolution spectroscopic study of BDs. We briefly introduce the binary BD system in \S \ref{sec:J0746AB}. We present the observations and data reduction in \S \ref{sec:observation} and \S \ref{sec:data_reduction}, respectively. In \S \ref{sec:data_analysis}, we describe our procedures for analyzing the reduced data. In \S \ref{sec:result}, we report results of J0746 AB observations. To interpret the results, we conduct simulations and a series of tests of the data analysis procedures. We also discuss the implications of the reported results. A summary and discussion are given in \S \ref{sec:summary}.

\section{2MASSW J0746425+200032AB (J0746 AB)}
\label{sec:J0746AB}

J0746 AB is a binary BDs at a distance of 12.21 pc. Because of its proximity to Earth, J0746 has been studied by multiple groups. Table \ref{tab:J0746} summarizes its properties from previous photometric, spectroscopic and radio observations.
~\citet{Bouy2004} estimated the spectral types for A and B (L0 and L1.5) and for the first time measured the dynamical mass of the system, which was later revised by~\citet{Konopacky2010} based on observations with a longer time baseline. Measurement of V$\sin i$~\citep{Blake2010, Konopacky2012} and inferred rotational period from photometric~\citep{Harding2013} and radio~\citep{Berger2009} monitoring indicated orbits aligned with the BD rotation~\citep{Harding2013}. No significant RV scattering was observed in addition to the orbital RV of A and B~\citep{Blake2010, Konopacky2010}. 

\section{Keck NIRSPEC Observation of J0746 AB}
\label{sec:observation}

We observed J0746 AB on UT 2016 Jan 26 at Keck II with the NIRSPEC high-resolution, infrared spectrograph~\citep{McLean1998, McLean2000} in seeing-limited mode. We used the NIRSPEC-7 filter, which results in partial wavelength coverage from 2.06 to 2.41 $\mu$m, spanning 5 echelle orders with gaps of $\sim$0.04 $\mu$m between orders. We used a slit length of 24$^{\prime\prime}$ and a slit width of 0.432$^{\prime\prime}$, corresponding to a spectral resolving power of $\sim24,000$. We adopted an ABBA nod pattern; an exposure time of 100 s was used for the first 16 exposures and 200 s long integrations were used for the subsequent 172 exposures at each position on the slit. We typically obtained 100--400 ADU (gain$=$5.8 e$^-$/ADU) per pixel depending on the airmass of the target. We observed the A1V star HIP 40058 ($K=6.1$) for telluric line correction.  Arc lamp lines were used for an initial wavelength solution; more precise wavelength calibration was achieved by fitting telluric absorption lines (\S5). Because observations were seeing-limited, the spectra of the A and B components in this 0.12$^{\prime\prime}$ pair are blended.

For over eight hours, we observed J0746 AB continuously with interruptions only for the telluric standard star observations, taken for every change of 0.2 in airmass or every hour, whichever came first. Seeing conditions improved from 1$^{\prime\prime}$ at the beginning of the night to 0.5$^{\prime\prime}$ at the end of the night. 

\section{Data Reduction}
\label{sec:data_reduction}

We reduced raw data from NIRSEPC using two packages, the IDL-based REDSPEC~\citep{Kim2015, Prato2015} and the Python-based PyNIRSPEC~\citep{Boogert2002, Piskorz2016}. We compared results from the two packages and concluded that the final, calibrated 1-d spectra are similar within measurement uncertainties. In the following, we describe the data reduction procedures of REDSPEC and PyNIRSPEC. 

\subsection{REDSPEC}
\label{sec:redspec}

Individual spectral pairs (AB or BA) were reduced using the Keck facility REDSPEC package\footnote{http://www2.keck.hawaii.edu/inst/nirspec/redspec/index.html}~\citep{Kim2015}, designed for extraction of single order NIRSPEC spectra. Ne, Ar, Xe, and Kr comparison lamp exposures provided initial dispersion and wavelength fiduciary point solutions. Sets of 10 flat fields and dark frames were taken and median filtered to create a master flat and dark for the removal of pixel inhomogeneities and dark current.

The stability of the echelle grating position was tested for every file across the $\sim$8 hours of observations by fitting a Gaussian to OH night sky lines to measure the relative variability of the centroid.  The maximum relative deviation in centroid value was $\sim$0.1 pixel or 0.4 km $\rm{s}^{-1}$.  Given the 0.432$^{\prime\prime}$ slit width, corresponding to $\sim$12 km $\rm{s}^{-1}$ velocity resolution, this represents grating stability at a 3\% level.

A and B frames were subtracted and divided by a dark-subtracted master flat.  The code then performed spatial and spectral rectification to account for the non-linear geometry of the data on the detector, resulting from the optimization of NIRSPEC's efficiency by use of the echelle grating in Quasi-Littrow mode.  The rectification was applied to the spectral traces of the NIRSPEC order spanning 2.309$-$2.342 $\mu$m, selected for our analysis of J0746 AB because of overlap with the spectral region observed by~\citet{Crossfield2014}.  Distortions in the spatial direction were fit with a second order polynomial; in the spectral direction, comparison lamp lines were fit in the cross-dispersion direction with second order polynomials, and their distribution across all columns in a given row with another second order fit.  

The REDSPEC code has two algorithms for the rectification procedure, one which simply interpolates over the original image and another which applies pixel mapping in order to accurately conserve the flux distribution in the pixels.  Although the difference in algorithms is only at the 1\% level, we choose to use the pixel mapping to minimize the introduction of any spurious artifacts into the data.  

After rectification, the two-dimensional spectra from one A$-$B pair were aligned with array rows in the dispersion direction.  Spectral rectification assigns equal wavelength increments to each pixel.
Spectra were extracted by summing the 8 rows over which 95\% of the flux was distributed.  Both a positive (corresponding to the A frame) and a negative (corresponding to the B frame) spectrum were extracted; these were then differenced a second time, eliminating any artifacts such as OH night sky emission lines.  We calculated the barycentric velocity correction for the center point of each extracted pair and corrected the pairs individually for this motion.

\subsection{PyNIRSPEC}
\label{sec:pynirspec}

Raw images were dark subtracted and flat fielded. A spectrum for each order was traced and straightened using a 3rd-order polynomial function. Collapsing all columns in the dispersion direction resulted in the creation of a PSF for the target. The PSF was then used for optimal extraction for 1-d spectrum~\citep{Horne1986}. 

Telluric lines were used for wavelength calibration. By comparing with synthetic telluric absorption spectra generated from the HITRAN database~\citep{Rothman2009}, we identified lines in 1-d spectrum. A wavelength solution was obtained by fitting centroid wavelengths of identified lines as a function of pixel with a 5th-order polynomial function. We chose 4 out of 5 echelle orders for subsequent data analysis because the four orders have higher SNR and more lines from the Earth's atmosphere and from the target. The wavelength ranges for the 4 orders were 2.38-2.42 $\mu$m (denoted as order 1), 2.31-2.34 $\mu$m (order 2), 2.24-2.28 $\mu$m (order 3), and 2.18-2.21 $\mu$m (order 4). 

Four exposures from each ABBA dither pattern resulted in one reduced spectrum. When combining spectra from the A and B positions, we considered weights that are inversely proportional to the square of flux uncertainty, which was calculated by the pipeline based on photon noise and detector noise. The combined spectrum was then normalized by a continuum function. The continuum is a 3rd-order polynomial fit to the median of each of 40 sections into which the spectrum was divided. 

\subsection{Comparing REDSPEC and PyNIRSPEC}
\label{sec:comp_red_py}

\begin{figure}
\epsscale{1.1}
\plotone{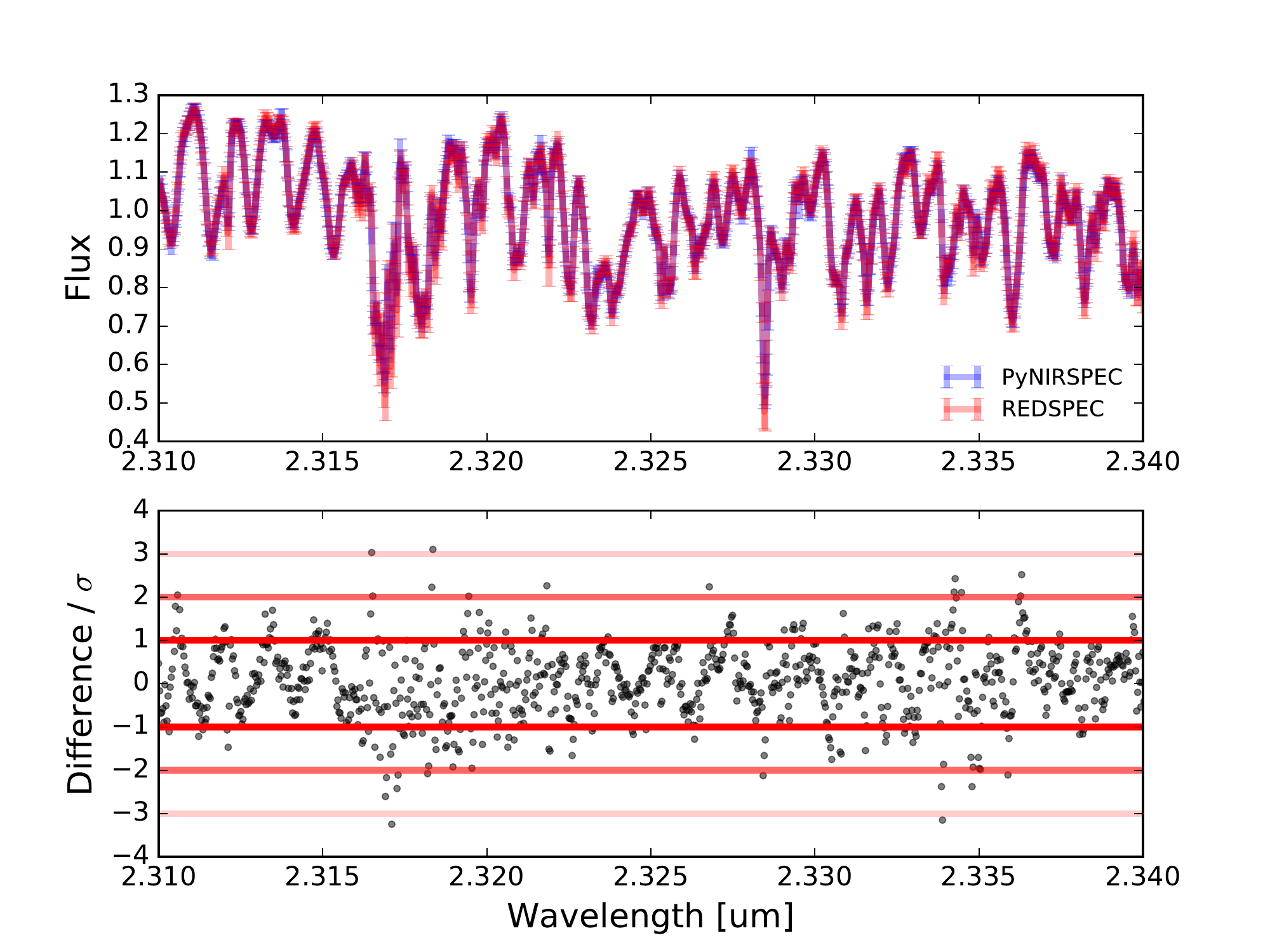}
\caption{Comparison of 1-d normalized wavelength-calibrated spectra from REDSPEC (red) and PyNIRSPEC (blue). Top: normalized flux as a function of wavelength with error bars. Bottom: the difference between REDSPEC and PyNIRSPEC spectra divided by combined uncertainty $\sigma$. Red horizontal lines with different shades indicate positive and negative 1$\sigma$, 2$\sigma$, and 3$\sigma$ lines. The two spectra agree with each other with a majority of the data points located within 1$\sigma$. 
\label{fig:spec_comp}}
\end{figure} 

We compared the spectra reduced using REDSPEC and PyNIRSPEC to determine if there was any systematic difference. Fig. \ref{fig:spec_comp} shows the comparison for the echelle order that cover 2.31-2.34 $\mu$m. The difference in the spectra shows structure and does not follow a normal distribution. This is attributed to slightly different wavelength solutions. However, the spectra reduced by two packages agree well within the uncertainty. The uncertainty is calculated by summing in quadrature the uncertainties from each package. PyNIRSPEC calculates the uncertainty based on photon noise and detector noise. REDSPEC does not provide an estimate of the uncertainty, we thus used 10 consecutive spectra and calculated the RMS error as an estimate of the uncertainty. Based on the comparison, PyNIRSPEC and REDSPEC result in similar reduced spectra. We used the spectra reduced by PyNIRSPEC in subsequent analyses. 

\section{Data Analysis}
\label{sec:data_analysis}

\subsection{Least Square Deconvolution}

We extracted line profile measurements with the least square deconvolution (LSD) technique~\citep{Kochukhov2010}. The spectral line profile was calculated from a linearly-sampled observed spectrum $\mathbf{Y^0}$ using the following Equation:
\begin{equation}
\label{eq:lsd}
\mathbf{Z=(M^T\cdot S^2\cdot M+\Lambda R)^{-1}\cdot M^T\cdot S^2\cdot Y^0}.
\end{equation}
In Equation \ref{eq:lsd}, the matrix transpose is denoted by $\mathbf{T}$. $\mathbf{M}$ is a $m \times n$ Toeplitz matrix, where $m$ is the number of data points in a spectrum and $n$ is the desired number of data points in the calculated line profile. $\mathbf{M}$ is generated from a template spectrum $\mathbf{F}$ that has the same wavelength sampling as the observed spectrum $\mathbf{Y^0}$. $\mathbf{S}$ is an $m \times m$ matrix with $\mathbf{S}_{ii} = 1 / \sigma_i$, where $\sigma_i$ is the measurement error for each spectral data point. The term $\mathbf{\Lambda R}$ is to regularize the LSD, where $\mathbf{\Lambda}$ is the regularization parameter and $\mathbf{R}$ is the matrix of first-order Tikhonov regularization. 

\subsection{Simultaneous Retrieval of Telluric and BD Line Profile}
\label{sec:sim_retrieval}

The observed spectrum consists of two components, telluric lines and spectral lines of BDs. The line profiles of the two components are independent and can be retrieved simultaneously. This requires an adaptation in Equation \ref{eq:lsd} as discussed in Section 2.4.1 in ~\citet{Kochukhov2010}. 

Specifically, two template spectra $\mathbf{F_1}$ and $\mathbf{F_2}$ are used instead of one, with $\mathbf{F_1}$ for the telluric spectrum and $\mathbf{F_2}$ for the BD spectrum. The template telluric spectrum is calculated based on the HITRAN database \citep{Rothman2009}. The BD template spectrum is obtained from the BT-Settl model\footnotetext{https://phoenix.ens-lyon.fr/Grids/BT-Settl/CIFIST2011\_2015/FITS/}~\citep{Baraffe2015}. We chose a spectrum withT$_{\rm{eff}}$ of 2300 K and log(g) of 5.0 as the BD template spectrum. Metallicity was set to solar abundance. Other metallicity options do not significantly affect the result. 

In this case, $\mathbf{F} = [\mathbf{F_1}, \mathbf{F_2}]$ is a $2n \times 1$ matrix. As a result, $\mathbf{M}$ is a $m \times 2n$ Toeplitz Matrix generated from $\mathbf{F}$. $\mathbf{\Lambda R}$ in Equation \ref{eq:lsd} becomes a matrix with two diagonal components, $\mathbf{\Lambda_1 R}$ and $\mathbf{\Lambda_2 R}$, where $\mathbf{\Lambda_1}$ is the regularization parameter for telluric line profiel retrieval, $\mathbf{\Lambda_2}$ is the regularization parameter for BD line profile retrieval, and $\mathbf{R}$ is again the matrix of first-order Tikhonov regularization.

\begin{figure*}
  \centering
  \begin{tabular}[b]{cc}
    \includegraphics[width=.5\linewidth]{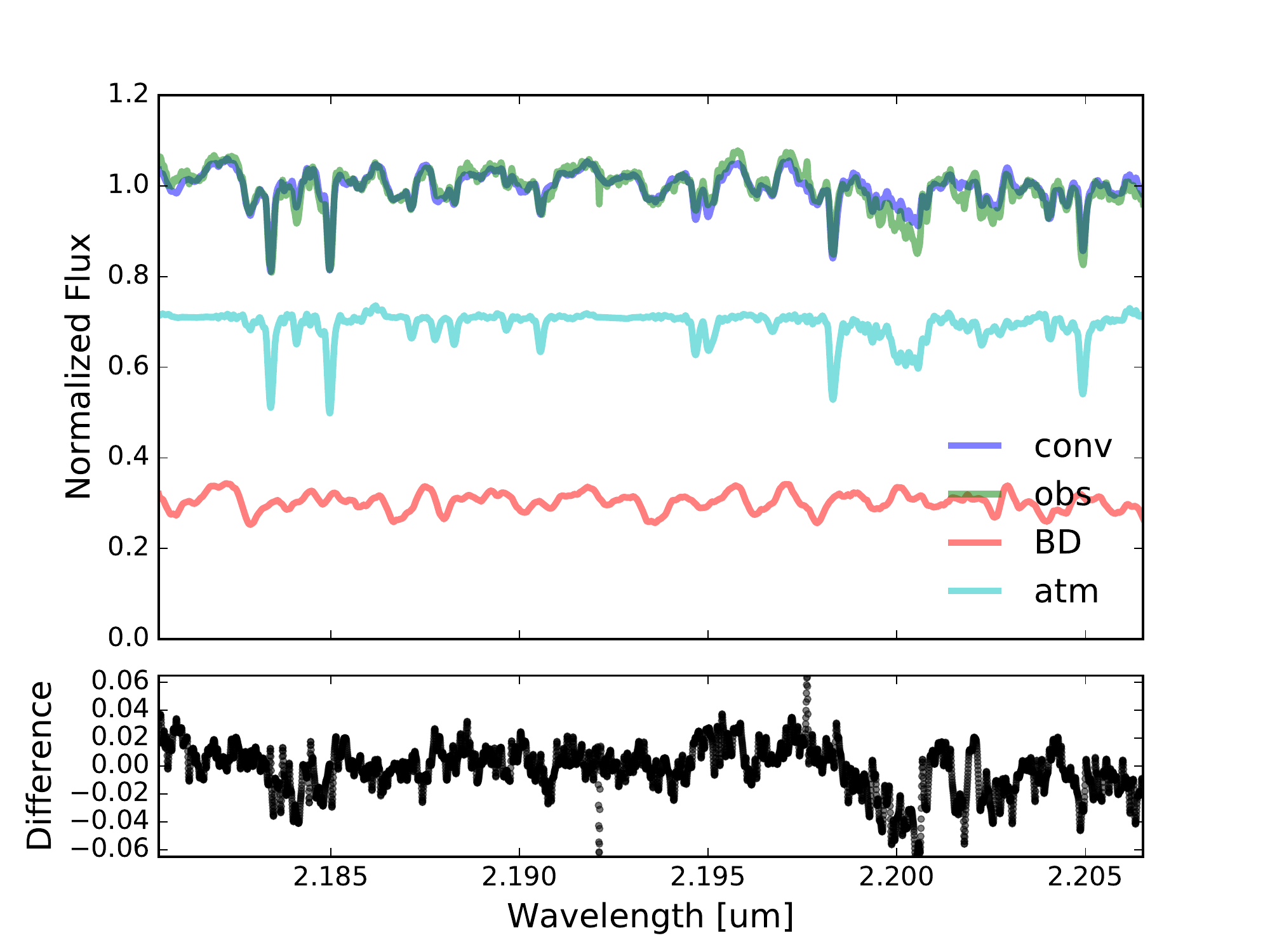} & \includegraphics[width=.5\linewidth]{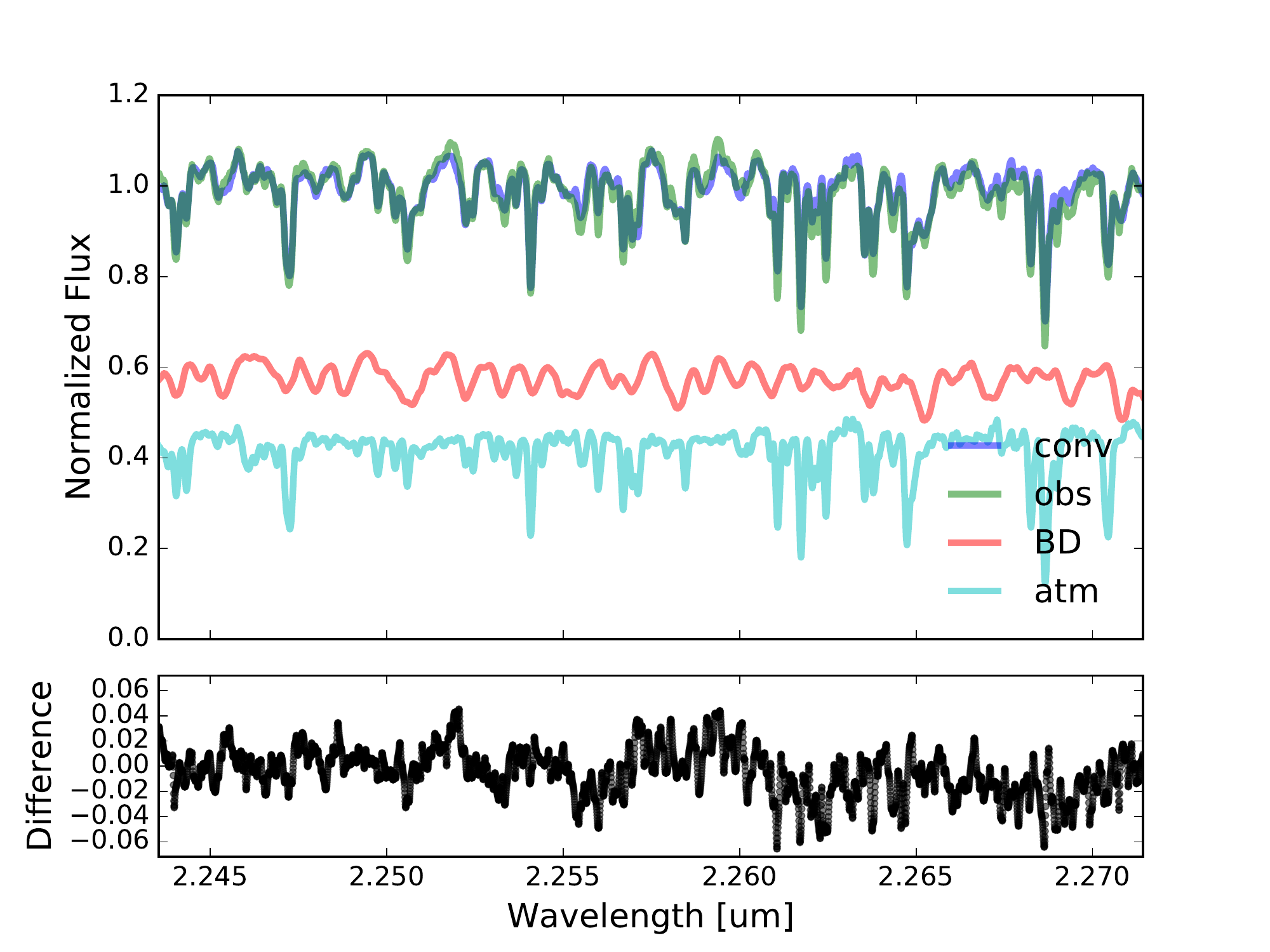} \\
    \includegraphics[width=.5\linewidth]{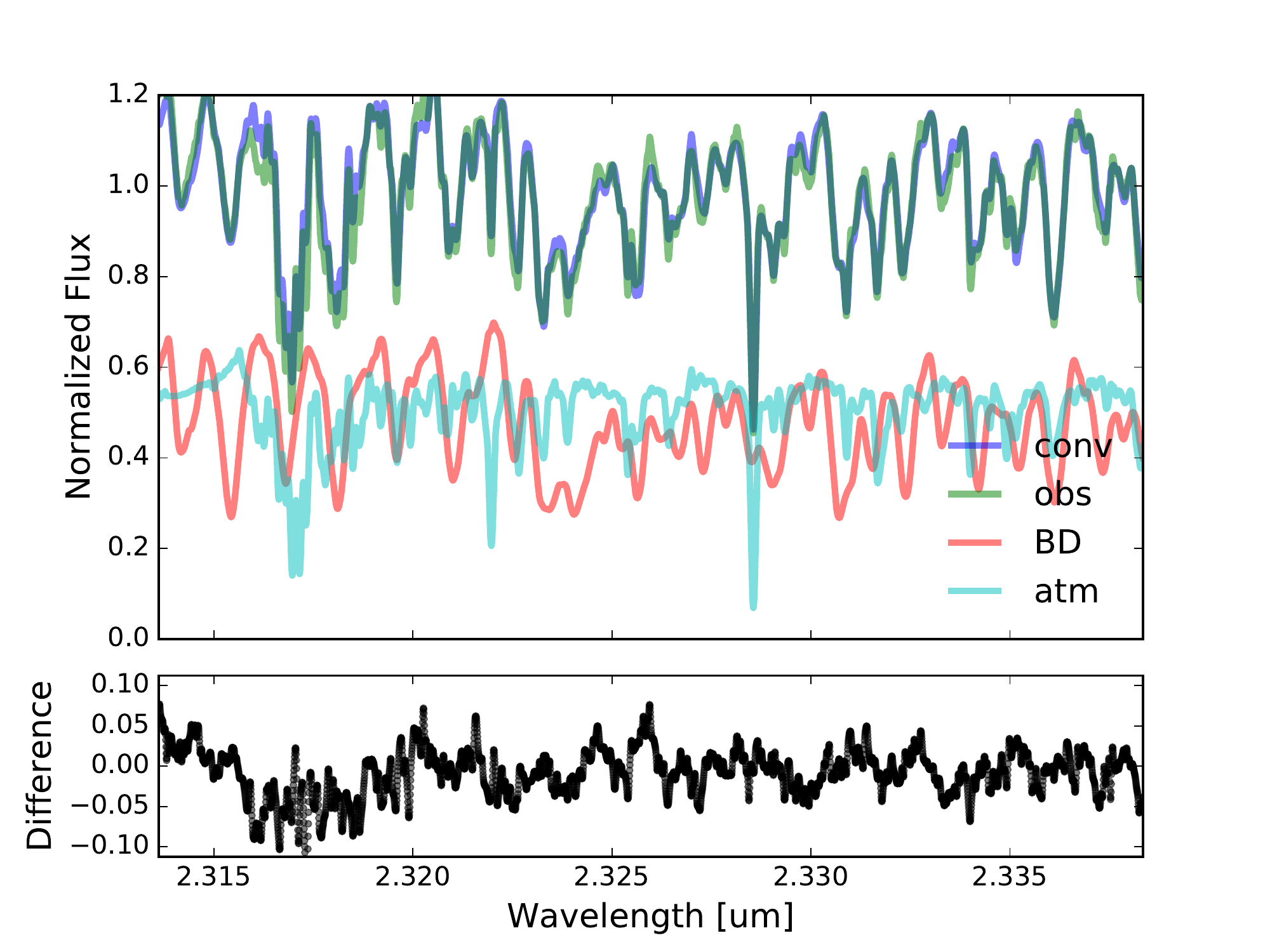} & \includegraphics[width=.5\linewidth]{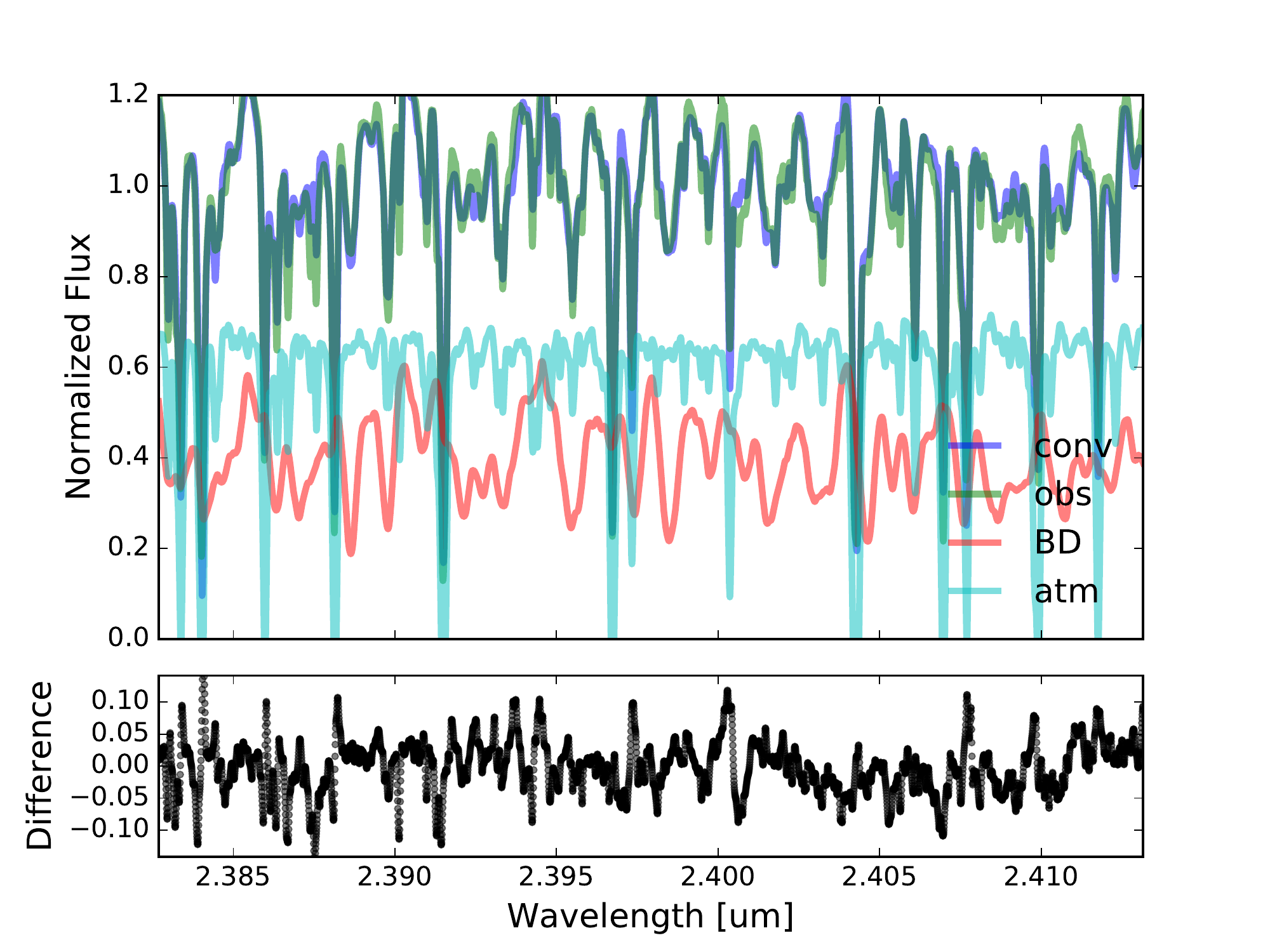} \\

  \end{tabular} \qquad
  \caption{Top panels: comparison of observed (green) and reconstructed (blue) spectra from retrieved line profiles for different orders. The reconstructed spectra have two components: telluric lines (cyan) and BD lines (red). Each component is calculated by convolving a template spectrum with the retrieved line profile from LSD. Bottom panels: the difference between the observed and reconstructed spectra. Typical RMS of the difference is 1.5-3.0\%.  \label{fig:lsd_order}}
\end{figure*} 

Fig. \ref{fig:lsd_order} shows the comparison of observed spectra (green) and the spectra reconstructed by retrieved line profiles (blue). The reconstruced spectra consist of two components, telluric spectrum (cyan) and BD spectrum (red). Each component is calculated by convolving a template spectrum with the retrieved line profile. The RMS is normally 1.5-3\% for the difference between the observation and the reconstructed spectrum, which is comparable with the typical uncertainty of 2\%. The difference can be attributed to photon-noise and the imperfect template used in LSD: a BD template has never been tested at a few percentage level at high spectral resolution.

\subsection{Correcting for Spectrograph Instability}

\begin{figure*}
  \centering
  \begin{tabular}[b]{ccc}
    \includegraphics[width=.3\linewidth]{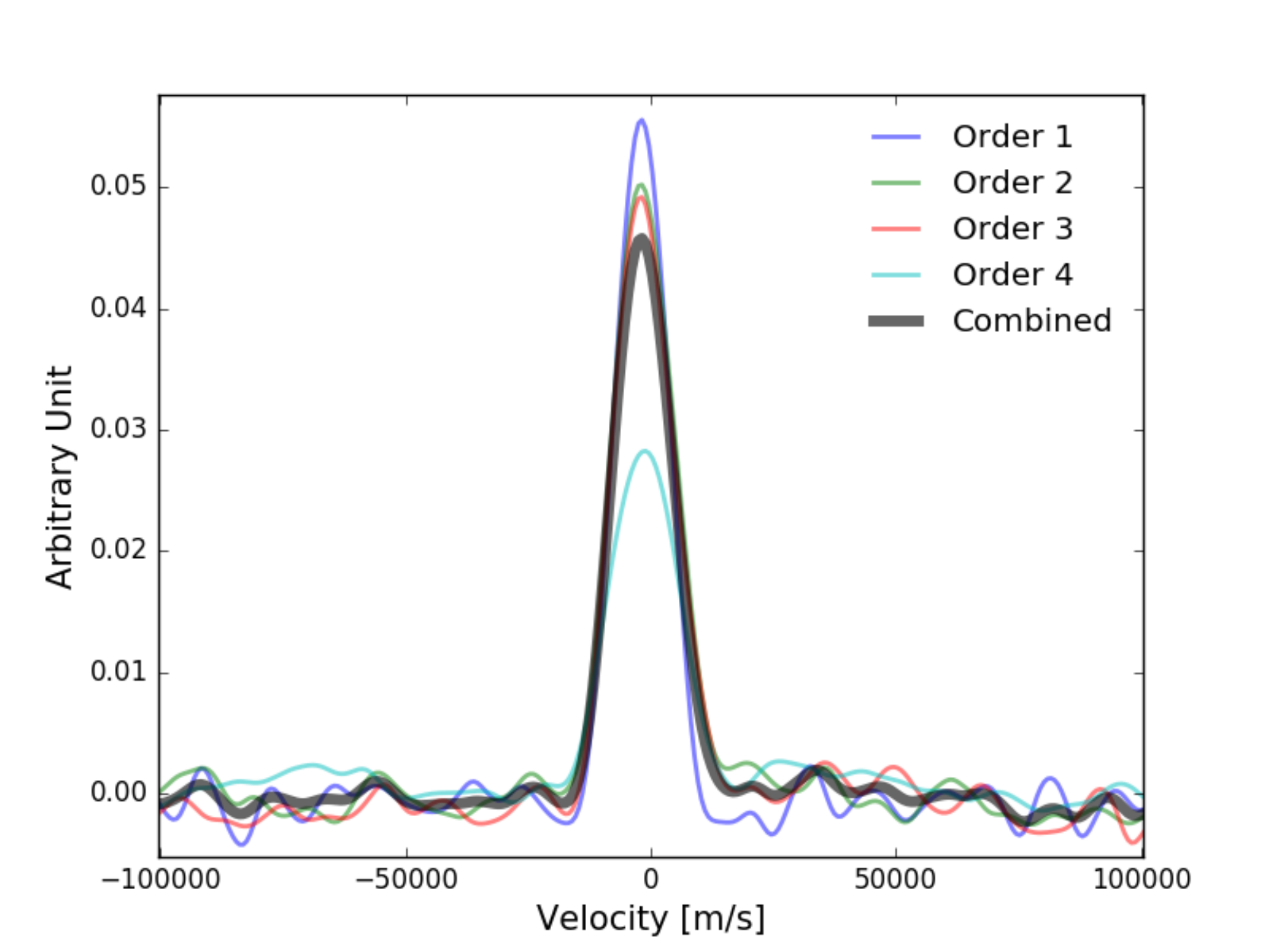} & \includegraphics[width=.3\linewidth]{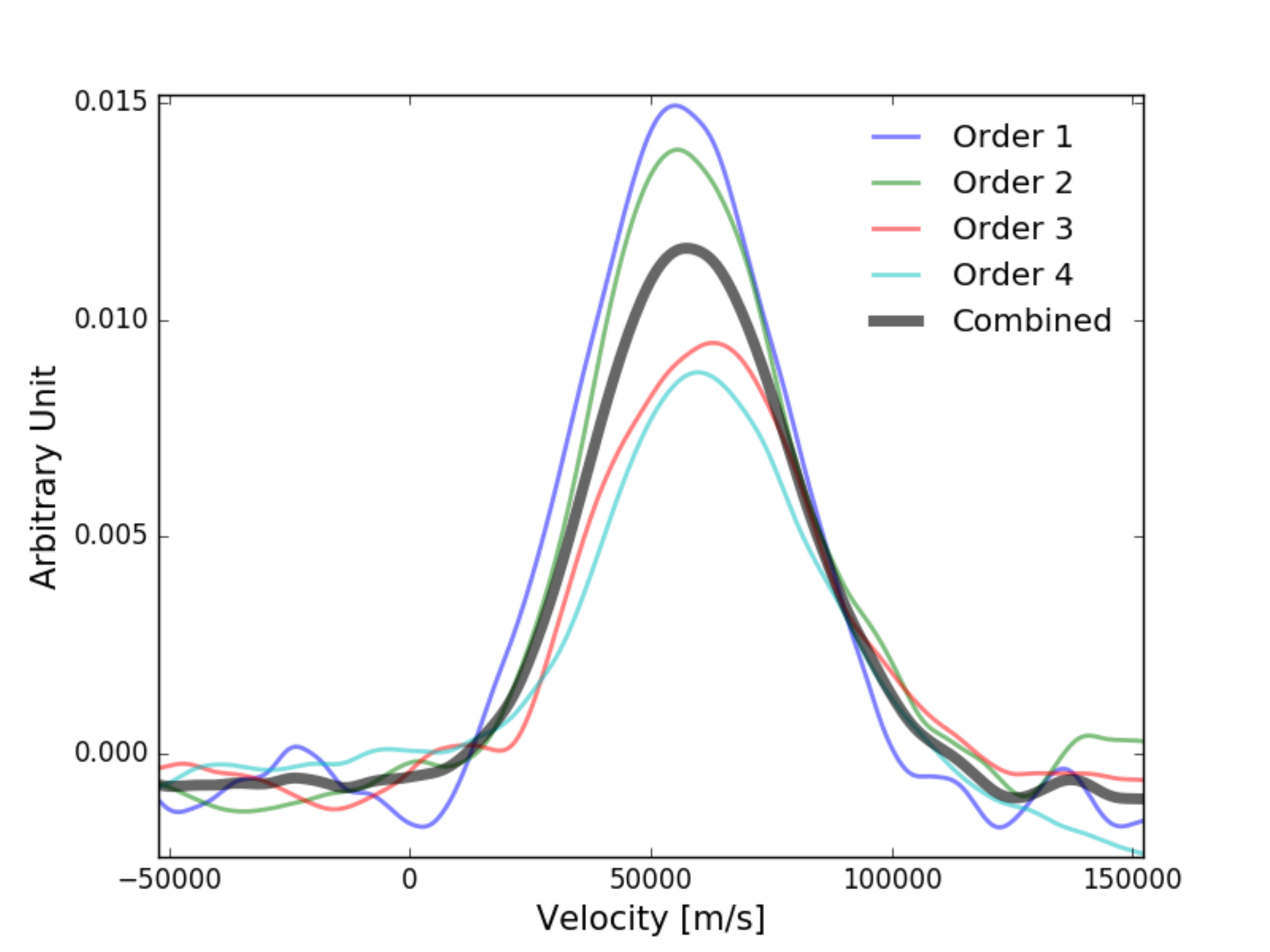} & \includegraphics[width=.3\linewidth]{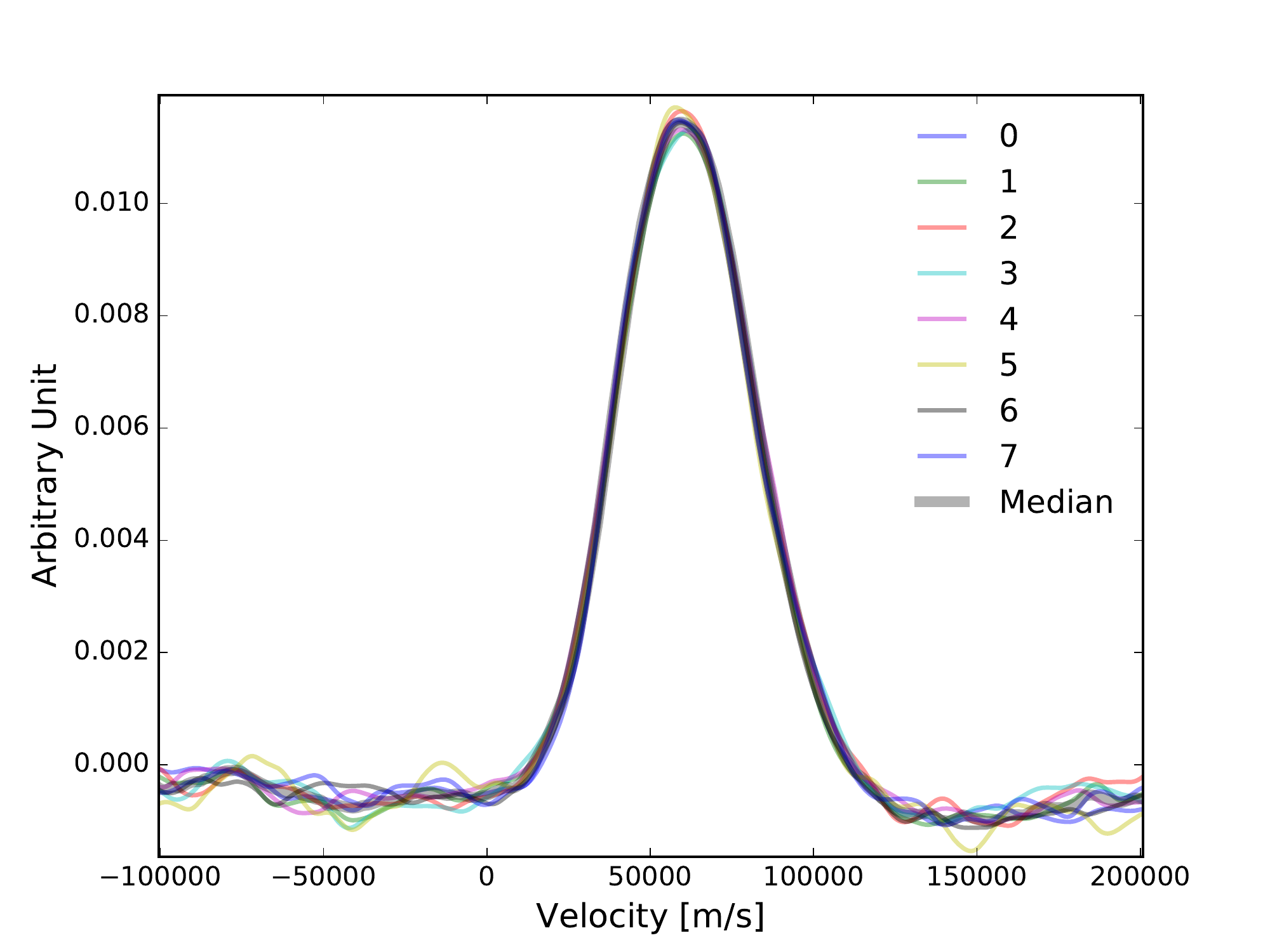} \\

  \end{tabular} \qquad
  \caption{Left: combined telluric line profile (black) from line profiles of different orders (colored). Middle: combined BD line profile (black) from line profiles of different orders (colored). Right: BD line profiles at different phases (colored) and the median line profile (black). \label{fig:illus_lsd}}
\end{figure*}

A spectrograph may be unstable during the observation, resulting in shifts of line profiles. We use telluric lines to correct for the line profile shift caused by instability in the spectrograph~\citep{Blake2010}. 

We first shifted and co-added telluric line profiles from the different orders (left in Fig. \ref{fig:illus_lsd}). The relative velocity shift between orders was recorded and was then used to shift and co-add BD line profiles (middle in Fig. \ref{fig:illus_lsd}). After obtaining the order-combined line profiles for the telluric and BD lines, we recorded the relative velocity shift between telluric line profiles for spectra taken at different times. The velocity shift was interpreted as the shift resulting from the spectrograph instability. We corrected for the instability by shifting the BD line profiles in the reverse direction with the same amount of velocity shift. After the correction, the BD line profiles were subtracted by the median line profile to find the deviation from the median line profile (right in Fig. \ref{fig:illus_lsd}). 

\section{Results}
\label{sec:result}

\subsection{Spectral Line Profile Variation for J0746 AB}
\label{sec:J0746_result}

\begin{figure*}
  \centering
  \begin{tabular}[b]{cc}
    \includegraphics[width=.4\linewidth]{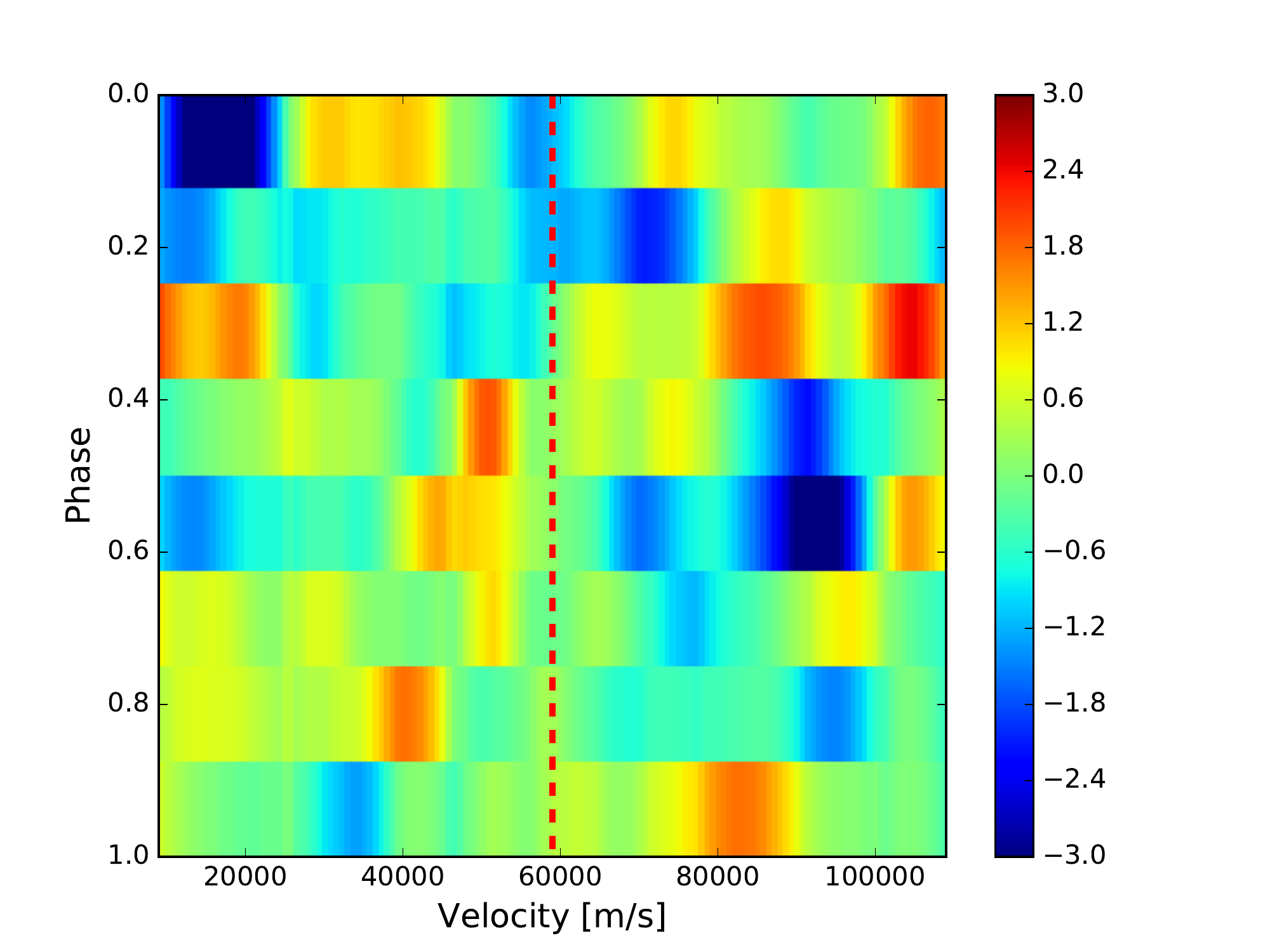} & \includegraphics[width=.4\linewidth]{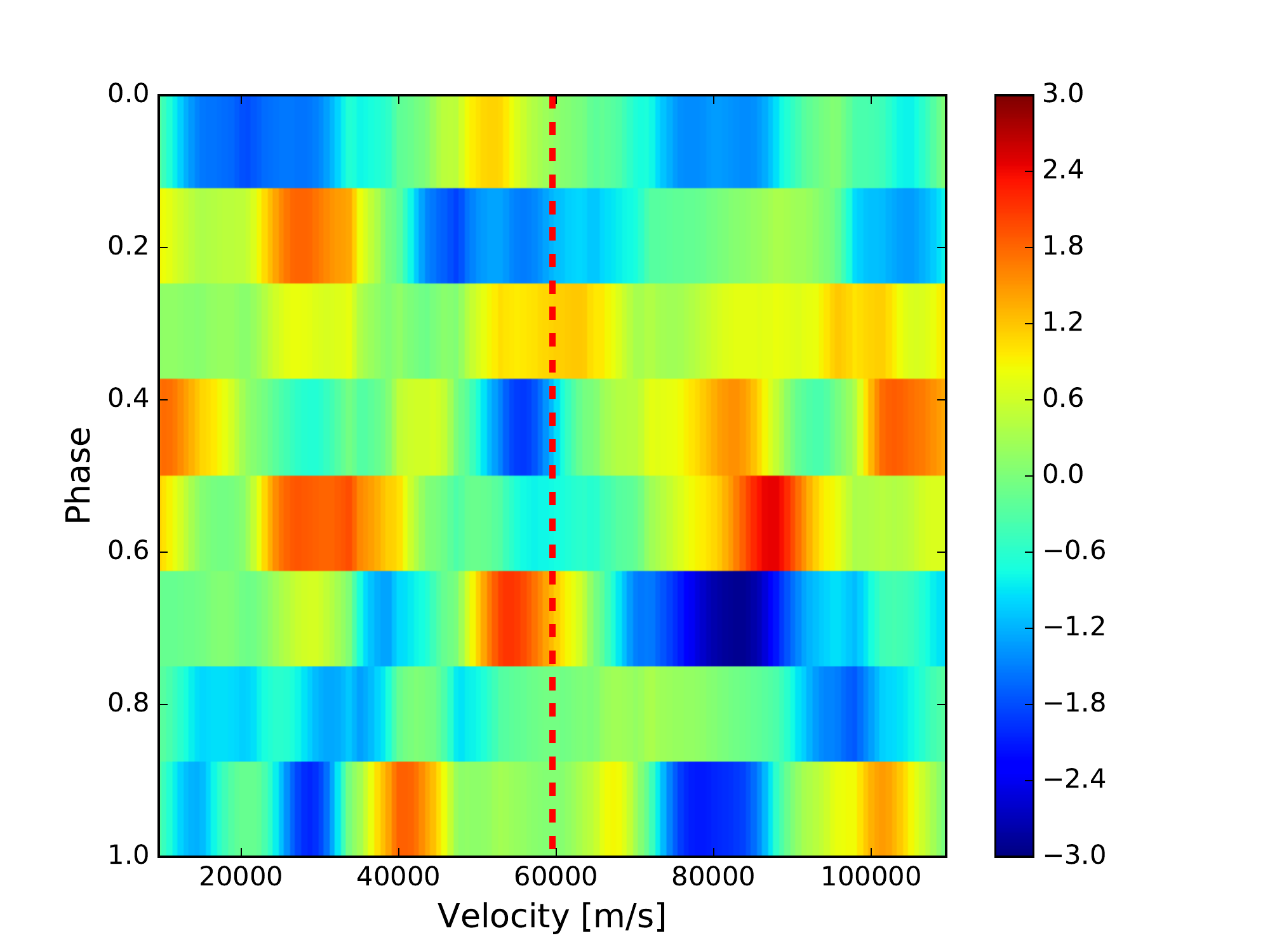} \\
  \end{tabular} \qquad
  \caption{Deviation plots for J0746 A (left) and B (right). These illustrate the measured deviation of line profiles from the median line profile as a function of rotation phase. The color scale was set so that red and blue correspond to plus and minus 3-$\sigma$ variations from the median value. The vertical red dashed line indicates the velocity difference between the BD and telluric lines.  \label{fig:J0746_result}}
\end{figure*}

The signal of J0746 A and B could not be separated spatially or spectrally in our data. The angular separation between A and B is 0.12$^{\prime\prime}$. Compared to a slit width of 0.432$^{\prime\prime}$ and the seeing conditions of 0.5-1.0$^{\prime\prime}$, A and B are unresolved and the spectra recorded on the detector contain signals from both components. In addition, The maximum RV difference between A and B is 8 km s$^{-1}$~\citep{Konopacky2010}. The projected rotational broadening (V$\sin i$) of A and B, 19 km s$^{-1}$ and 33 km s$^{-1}$, makes it difficult to separate signal from A and B in velocity space. 

J0746 A and B have different rotation periods; despite the blended signal, we can fold the observed spectra based on these periods. As a result, the signal that is modulated with rotation period is enhanced whereas the signal out of the periodicity is reduced. Therefore, we repeat the following procedure for both A and B in order to search for coherent variability that modulates with rotation. 

We divided one rotation period into 8 phases. Zero phase is the MJD time that is dividable by the rotation period. We selected spectra taken around each phase and then took the median of these spectra as the phase-folded spectrum. The uncertainty was the standard deviation of the spectra used to construct the phase-folded spectrum. There were typically 4-5 spectra to combine for each phase. This resulted in 8 spectra for 8 phases (see right panel of Fig. \ref{fig:illus_lsd}). 

\begin{figure*}
  \centering
  \begin{tabular}[b]{cc}
    \includegraphics[width=.4\linewidth]{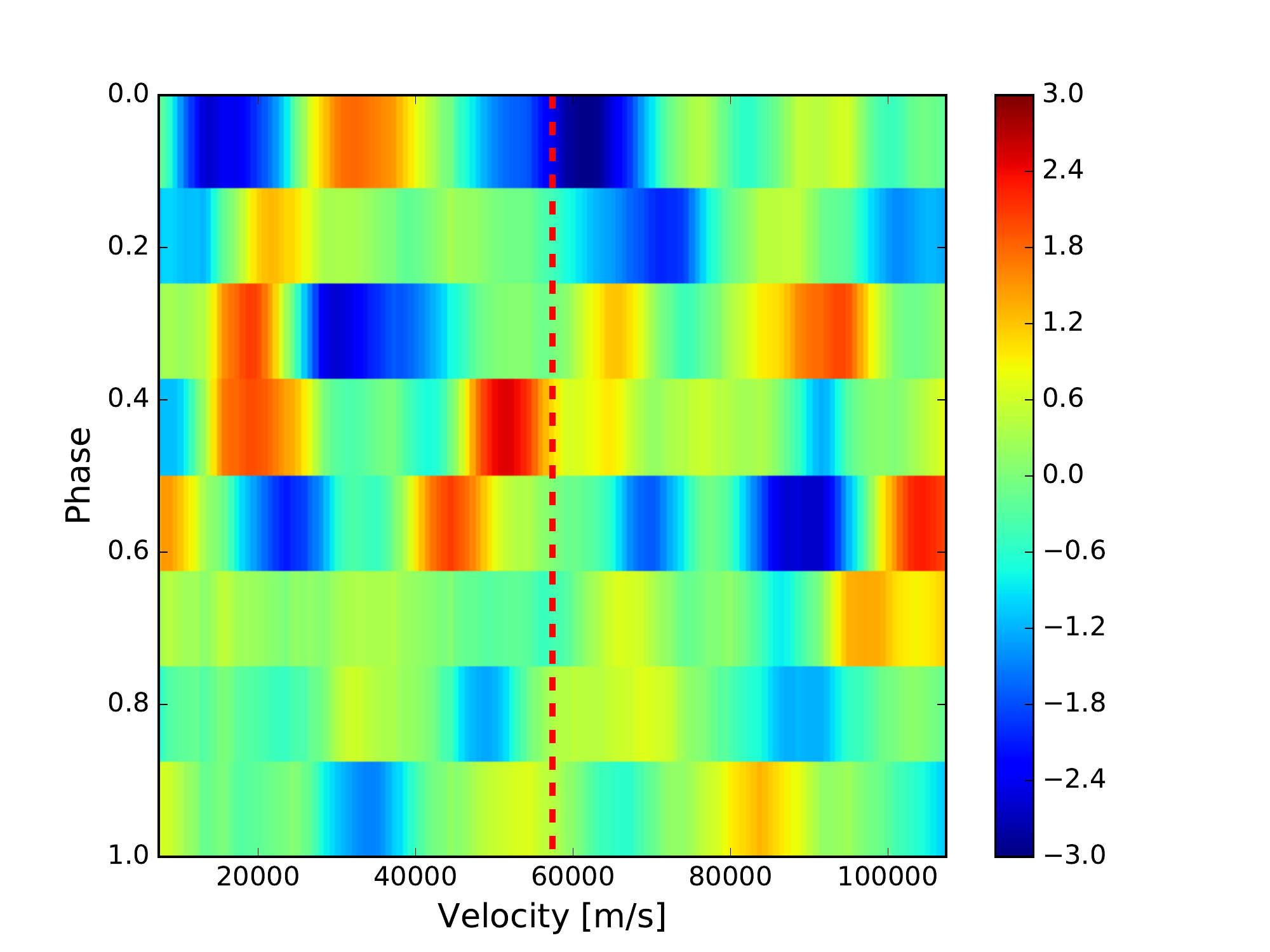} & \includegraphics[width=.4\linewidth]{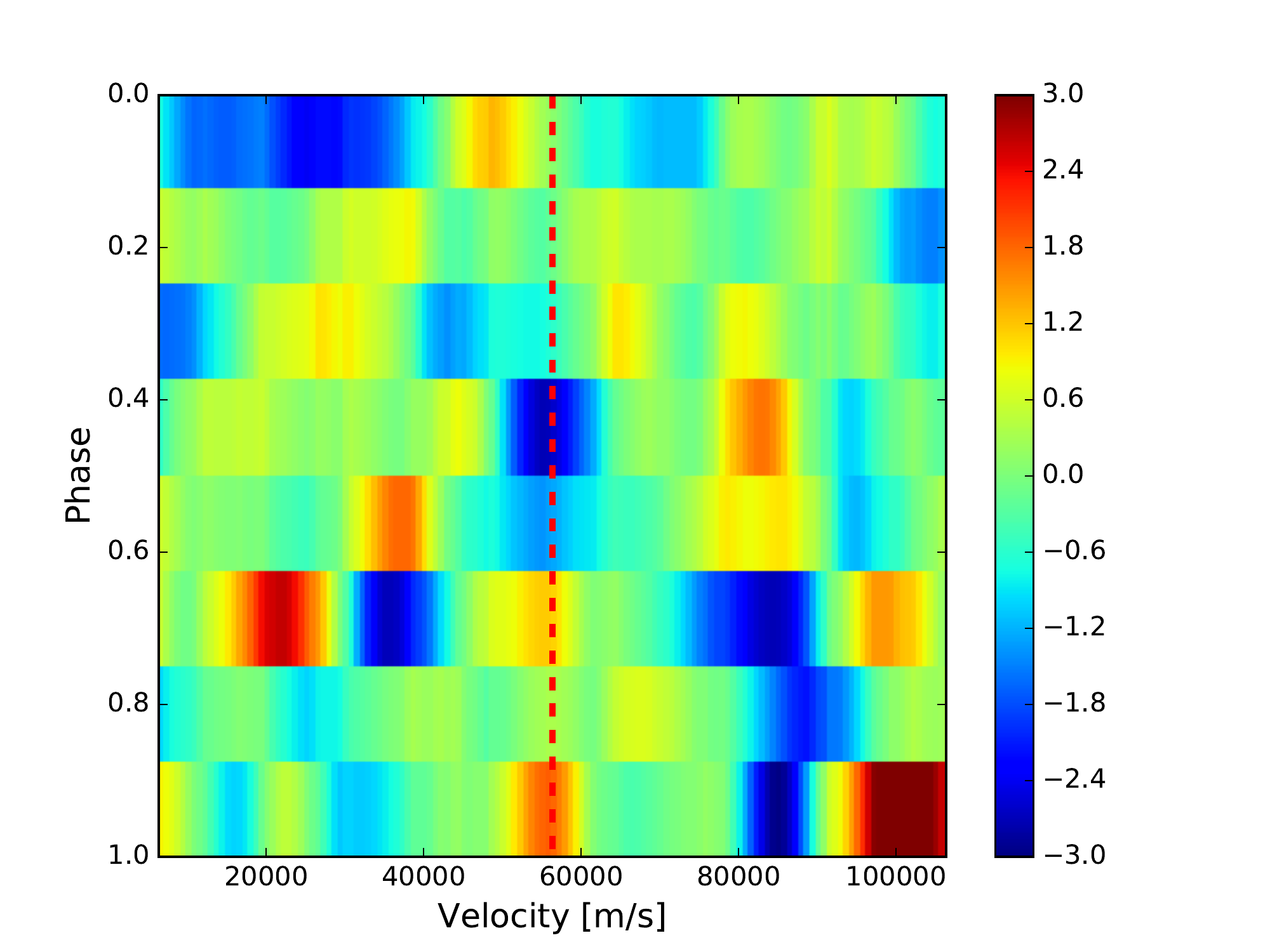} \\
    \includegraphics[width=.4\linewidth]{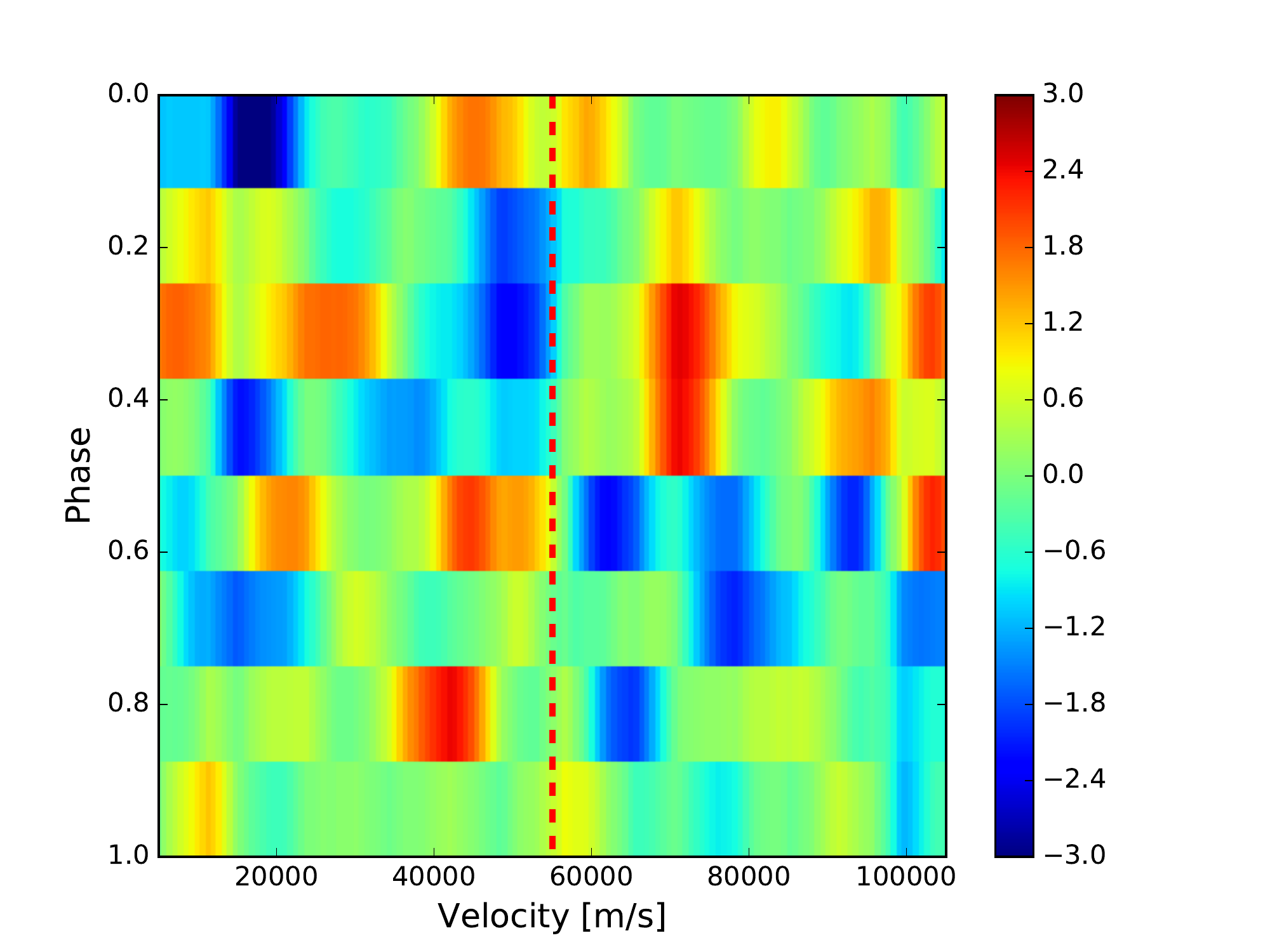} & \includegraphics[width=.4\linewidth]{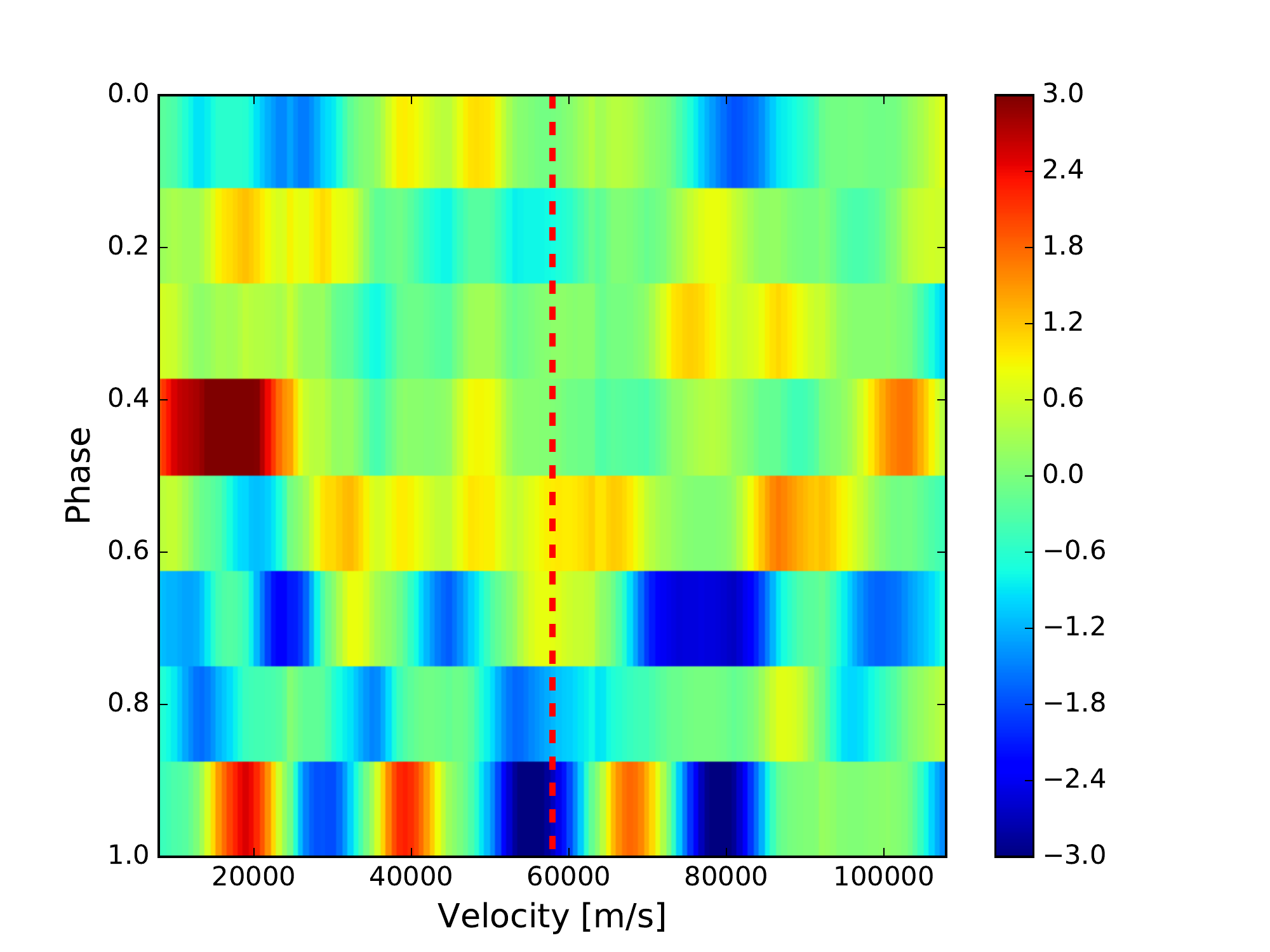} \\
    \includegraphics[width=.4\linewidth]{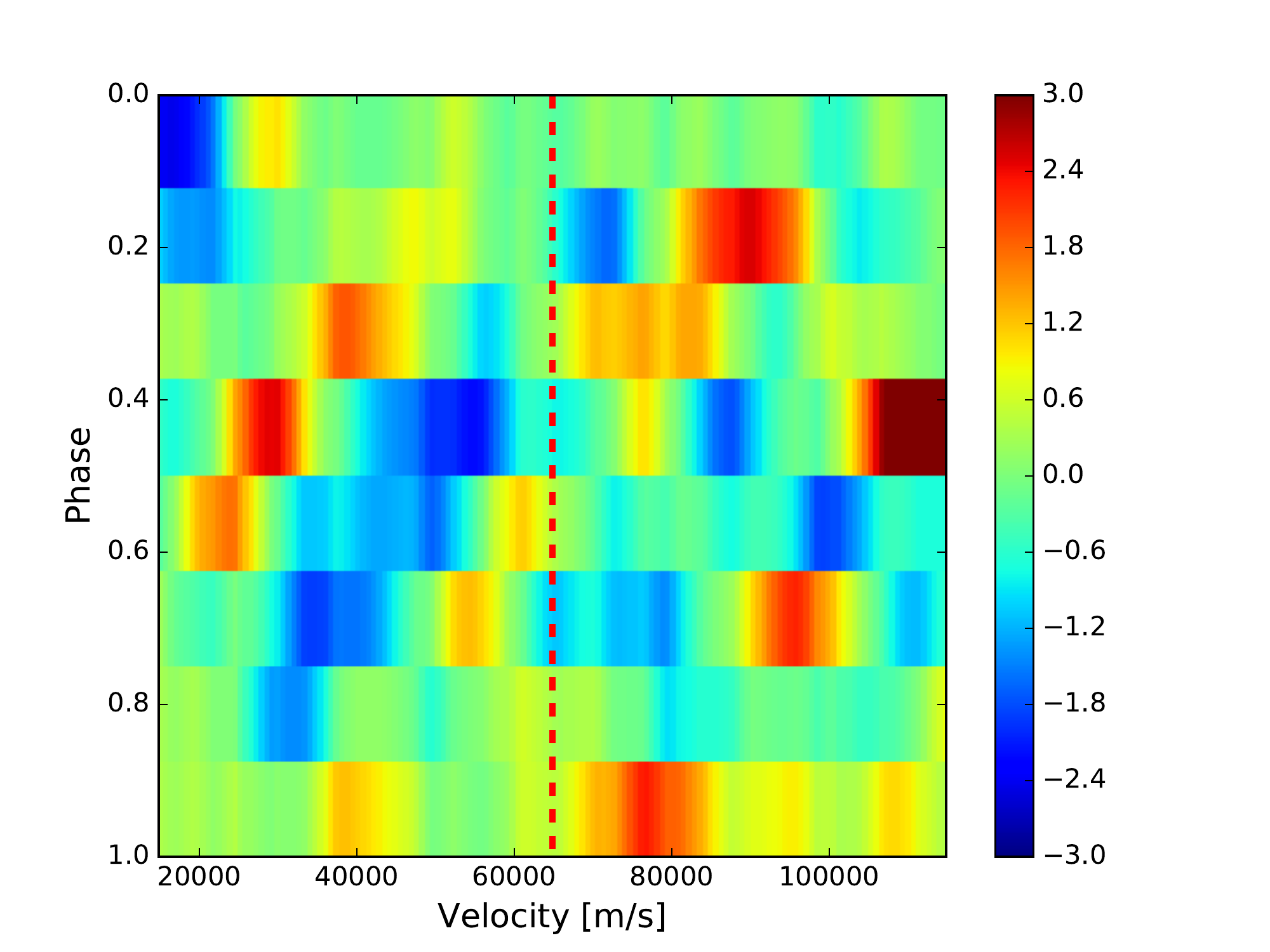} & \includegraphics[width=.4\linewidth]{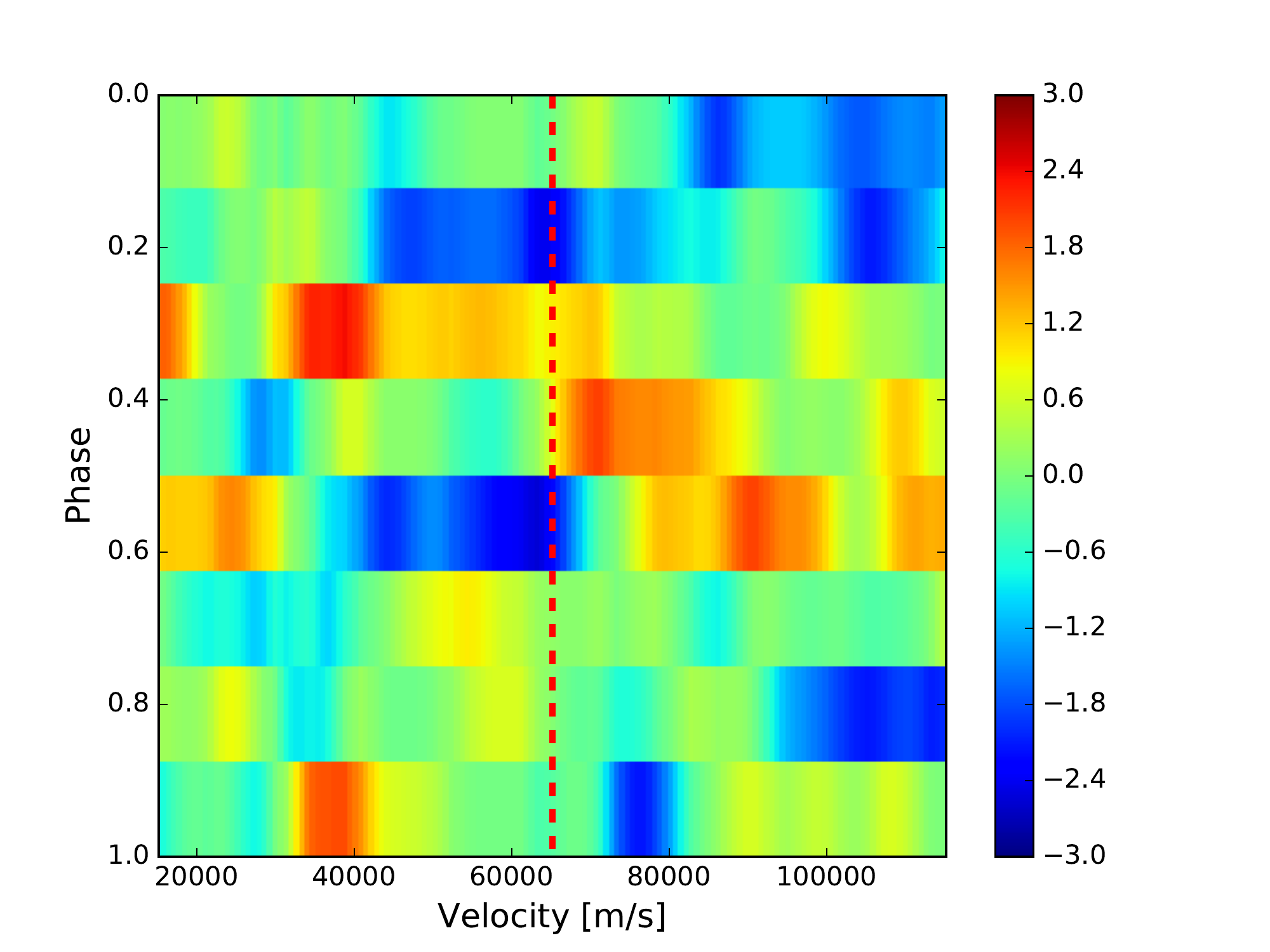} \\
    \includegraphics[width=.4\linewidth]{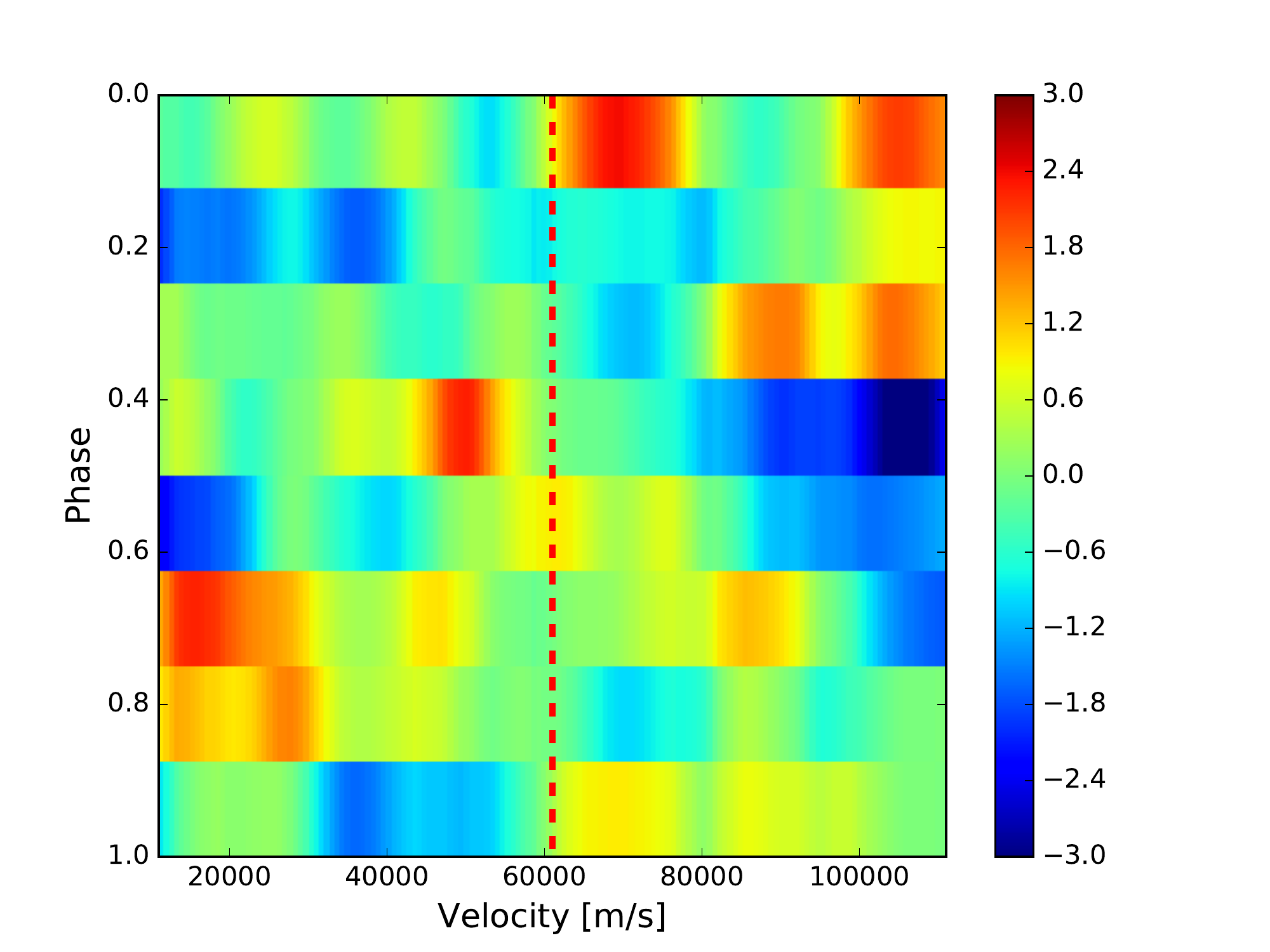} & \includegraphics[width=.4\linewidth]{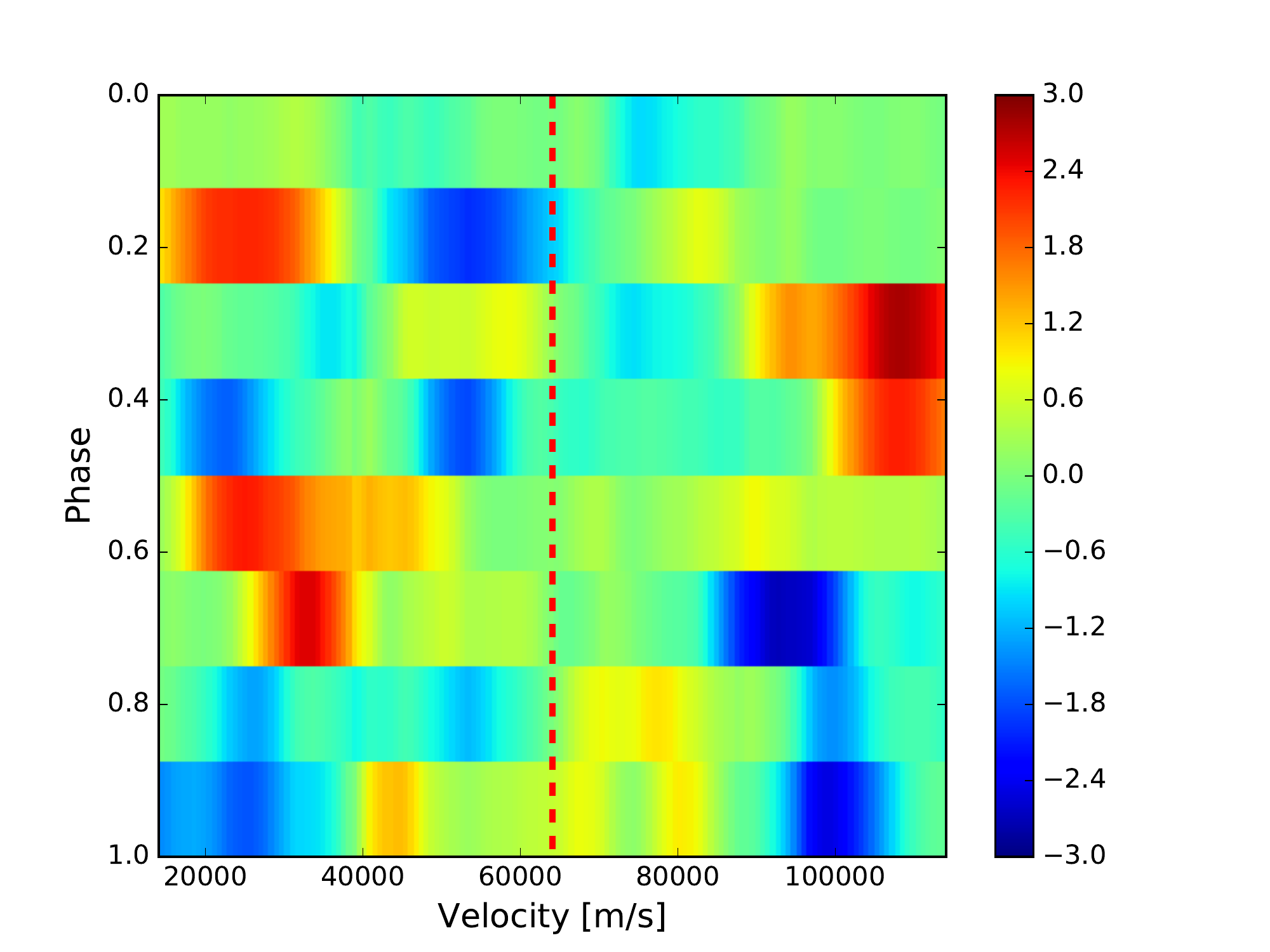} \\
  \end{tabular} \qquad
  \caption{Deviation plots for J0746 A (left) and B (right) for a single echelle order (order 1-4 from top to bottom). Refer to Fig. \ref{fig:lsd_order} for wavelength coverage for each order. Refer to left and middle panel in Fig. \ref{fig:illus_lsd} for relative signal contribution from each order. The color scale is set so that red and blue correspond to plus and minus 3-$\sigma$ variations from the median value. The vertical red dashed line indicates the velocity difference between the BD and telluric lines. \label{fig:J0746_result_by_order}}
\end{figure*}

Fig. \ref{fig:J0746_result} shows the deviation plot for A (left) and B (right). A deviation plot shows the difference between the line profile for a certain phase and the median line profile of all measurements. We looked for coherent deviation of line profile as a function of time similar to what~\citet{Crossfield2014} detected for Luhman 16 B. However, no coherent deviation was detected at 3-$\sigma$ level for either A and B. We used the standard deviation of the deviation plot as 1-$\sigma$. We also checked the deviation plots for different orders (Fig. \ref{fig:J0746_result_by_order}), but no consistent variation was detected. 

\subsection{Understanding the Non-detection}
\label{sec:test_sim}

The spectral variability of J0746 AB was not detectable for our NIRSPEC seeing-limited spectra and data reduction techniques. In order to understand the sensitivity of our observations and the implications of the non-detection, we simulated NIRSPEC observations and applied the data analysis procedure described in \S \ref{sec:data_analysis} to the simulated data. 

\subsubsection{Generating Spherically-Integrated Spectra}

We divided a unit sphere into a 200 by 100 grid in longitude ($\alpha$) and latitude ($\delta$). For each grid point, we calculated the angle $\mu$ between the normal surface and the line of sight. The radial velocity (RV) for each grid point was V$\times\cos(\delta)\times\sin\mu$ where V is the equatorial rotational velocity. The limb darkening $I(\cos\mu)/I(1)$ for each grid point was $1-a\times(1-\cos\mu)-b\times(1-\cos\mu)^2$, where $a=0.3699$ and $b=0.2483$ are parameters for quadratic darkening~\citep{Claret2012}. The surface area projected to line of sight $\Delta A$ was $\Delta\alpha\times\Delta\delta\times\cos\delta\times\cos\mu$, where $\Delta\alpha$ and $\Delta\delta$ represented angular size for each grid point. 

Each grid point was assigned a spectrum $\rm{F}(\lambda)$, where $\lambda$ denotes wavelength. We used the BT-Settl spectrum with T$_{\rm{eff}}$ of 2300 K, log(g) of 5.0 and [Fe/H]=0.0. We applied a Doppler shift ($\delta\lambda$) to the spectrum based on the RV for each grid point, and then added up spectra for all grid points to obtain the integrated spectrum $\rm{I}(\lambda)$ with the following equation: $$\rm{I}(\lambda) = \frac{\sum_{i=1}^{n}\rm{F}(\lambda+\delta\lambda_i)\times I(\cos\mu_i)/I(1)\times\Delta A_i}{\sum_{i=1}^{n} I(\cos\mu_i)/I(1)\times\Delta A_i}$$, where $i$ denotes grid point index and $n$ is the total number of grid points. 

For a sphere with non-uniform surface brightness $f_i$, the integrated spectrum can be calculated with the following equation:
\begin{equation}
\label{eq:integrated_spot_spec}
\rm{I}(\lambda) = \frac{\sum_{i=1}^{n}f_i\times\rm{F}(\lambda+\delta\lambda_i)\times I(\cos\mu_i)/I(1)\times\Delta A_i}{\sum_{i=1}^{n}f_i\times I(\cos\mu_i)/I(1)\times\Delta A_i}.
\end{equation}
With Equation \ref{eq:integrated_spot_spec}, we can generate an integrated spectrum for a target covered by spots given its surface brightness map $f_i$. 

We generated a series of surface maps with a spot on the equator of a revolving object for two rotational periods. We chose two periods because (1) our observation spanned about two rotation periods of A and (2) we wanted to demonstrate the repeatability of the coherent signal. We considered two cases for the spot contrast, 0.5 or 0.8 times the normal brightness, and variable spot size. The two contrast values are consistent with the upper and lower limit from observations~\citep{Crossfield2014, Karalidi2016}. The initial position of the spot was at anti-phase with respect to an observer, i.e., the observer and the spot are on opposite sides of the sphere. We matched the inclination and rotation velocity of the system to those of J0746. Based on Equation \ref{eq:integrated_spot_spec}, we generated a series of  integrated spectra as the spot revolved around the rotation axis. 

\subsubsection{Simulating Observations}
\label{sec:sim_obs}

To simulate observations, we multiplied grid-flux-integrated spectra by a telluric spectrum. The resulting spectra were then convolved with an instrumental profile. The instrumental profile was chosen to match the spectral resolution (R$\sim$25,000) with a FWHM of $\sim$12 km/s. We assumed the instrumental profile was stable. In a situation with an unstable instrumental profile, the position of telluric lines can be used to track instrument instability (see \S \ref{sec:data_analysis}). 


We added Gaussian noise ($\sigma=0.02$) to the noiseless simulated spectra. The standard deviation of the Gaussian noise was comparable to typical uncertainty per pixel of 2\%. We analyzed the spectra with the same procedures as outlined in \S \ref{sec:data_analysis}. This analysis of simulated data helps us to understand the sensitivity limit to surface non-uniformity at a given noise level. 

\subsubsection{Results on Simulated Data}

\begin{figure*}
  \centering
  \begin{tabular}[b]{cc}
    \includegraphics[width=.4\linewidth]{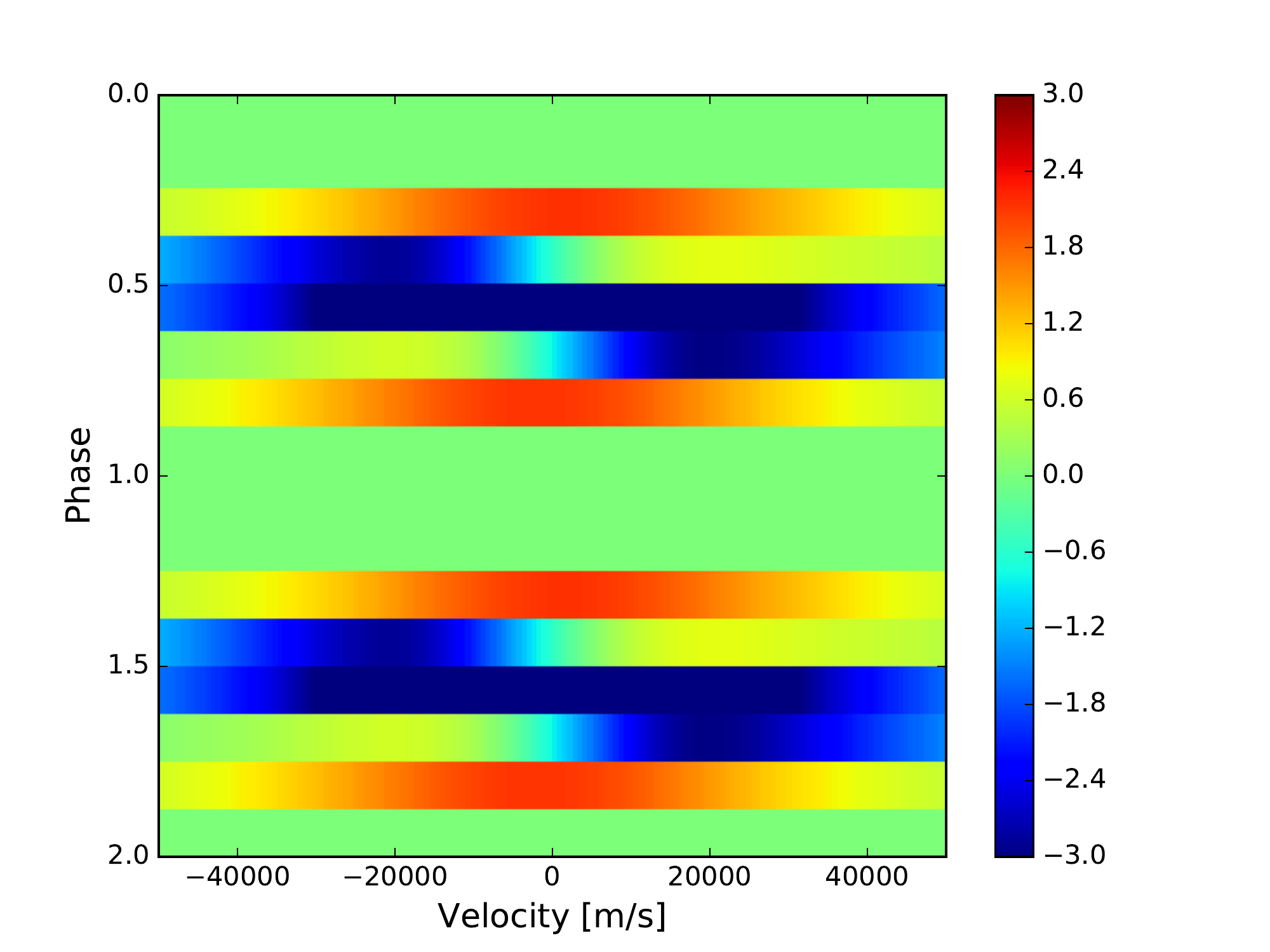} & \includegraphics[width=.4\linewidth]{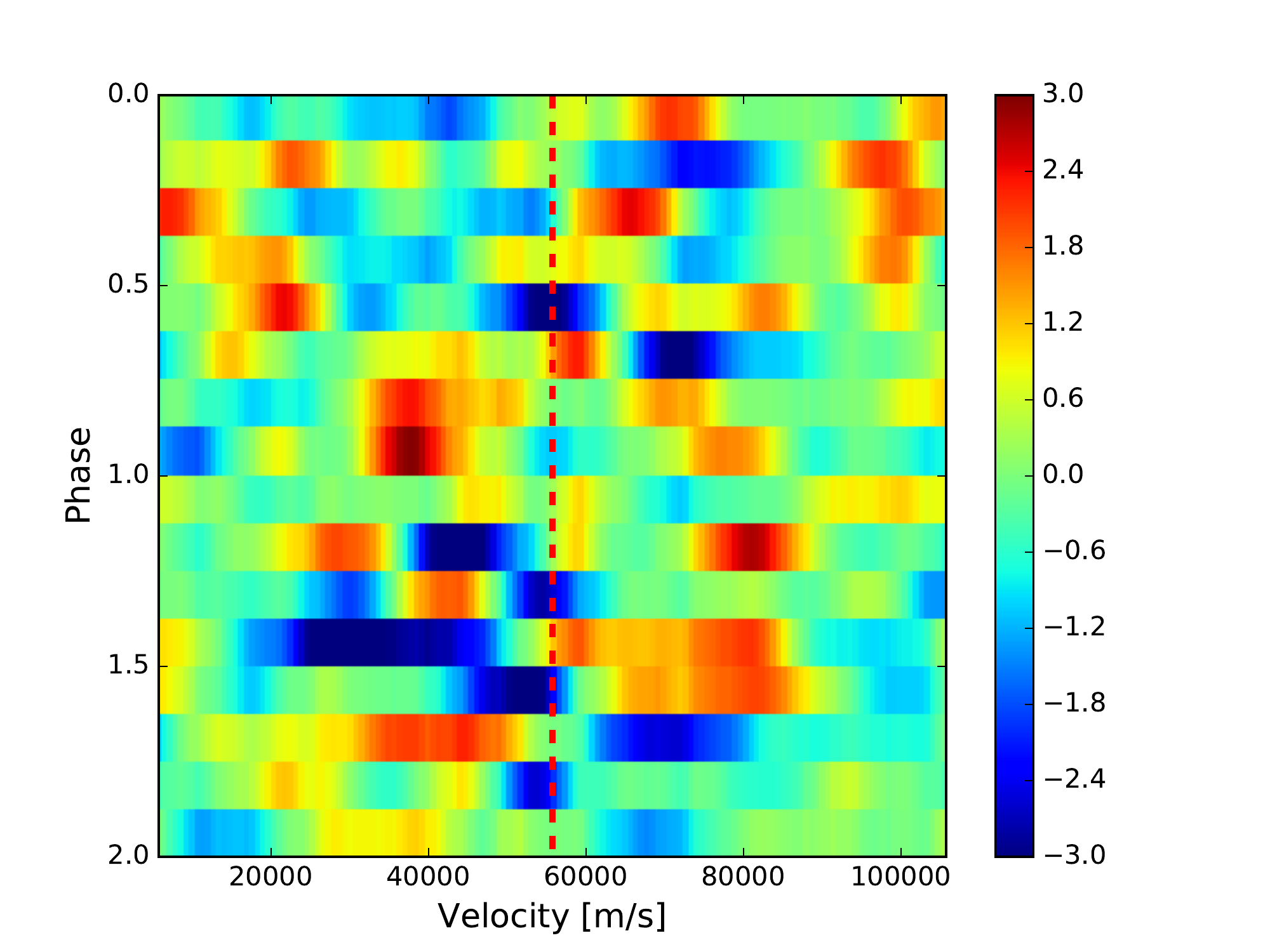} \\
    \includegraphics[width=.4\linewidth]{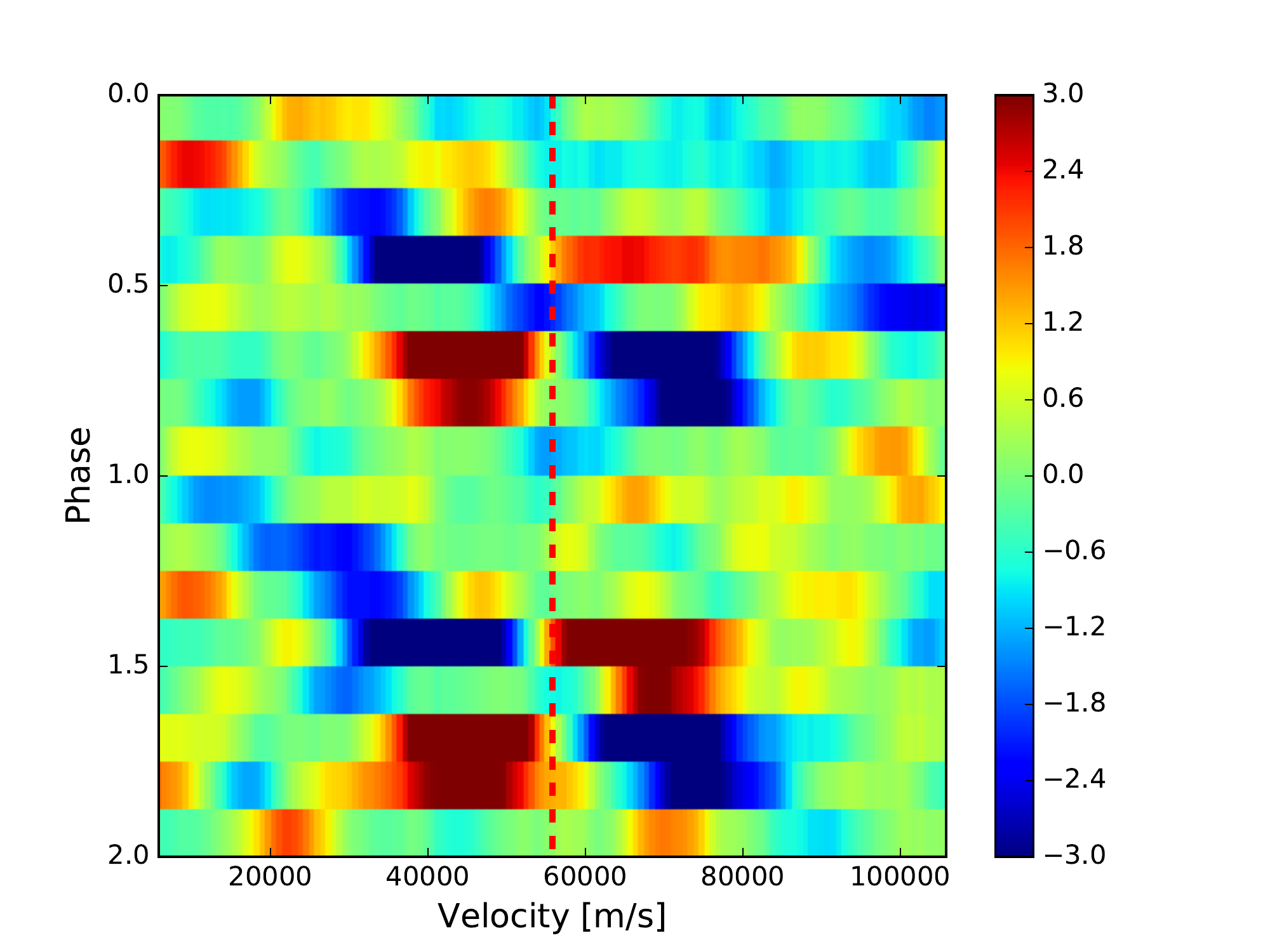} & \includegraphics[width=.4\linewidth]{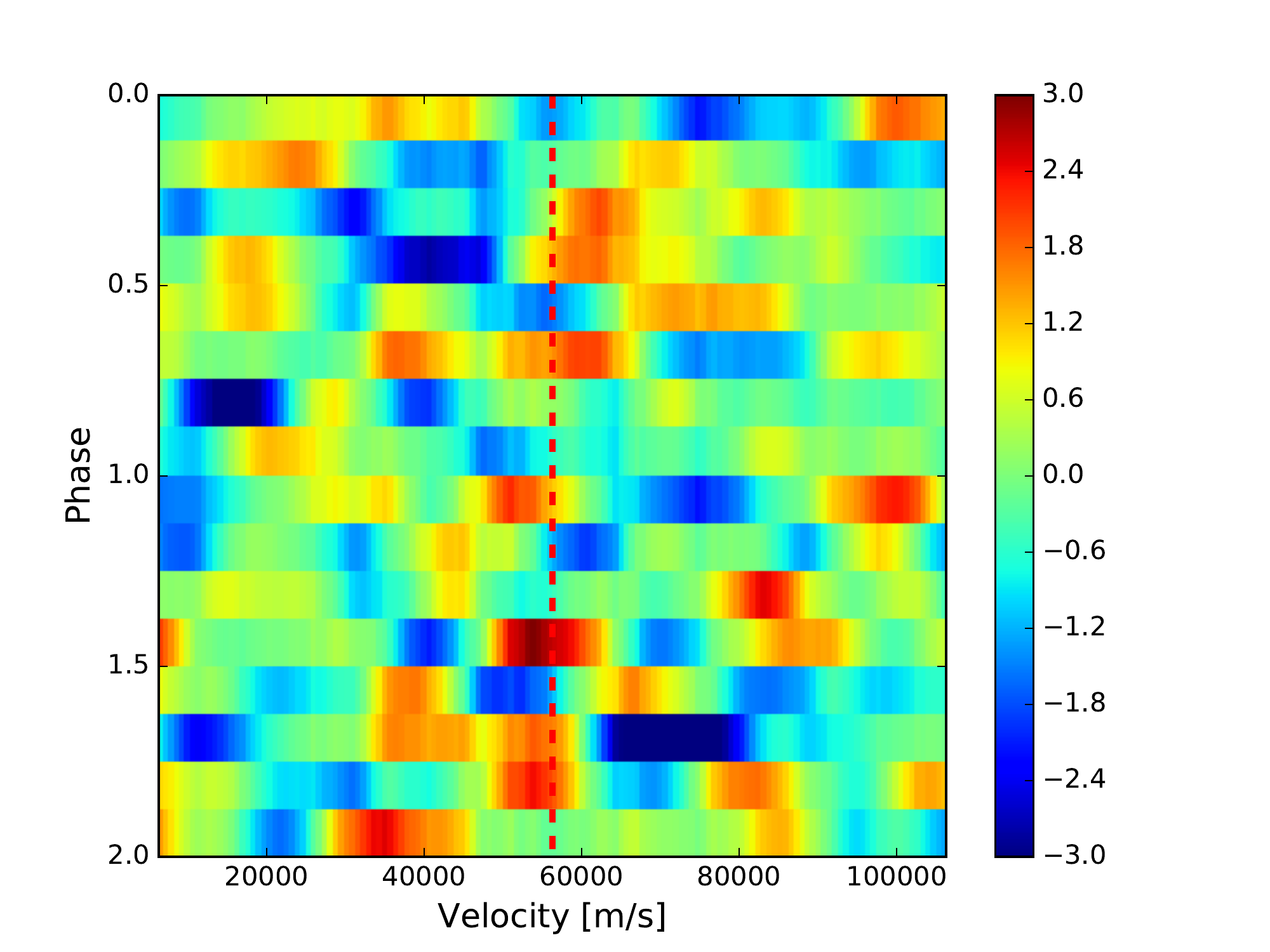} \\
    \includegraphics[width=.4\linewidth]{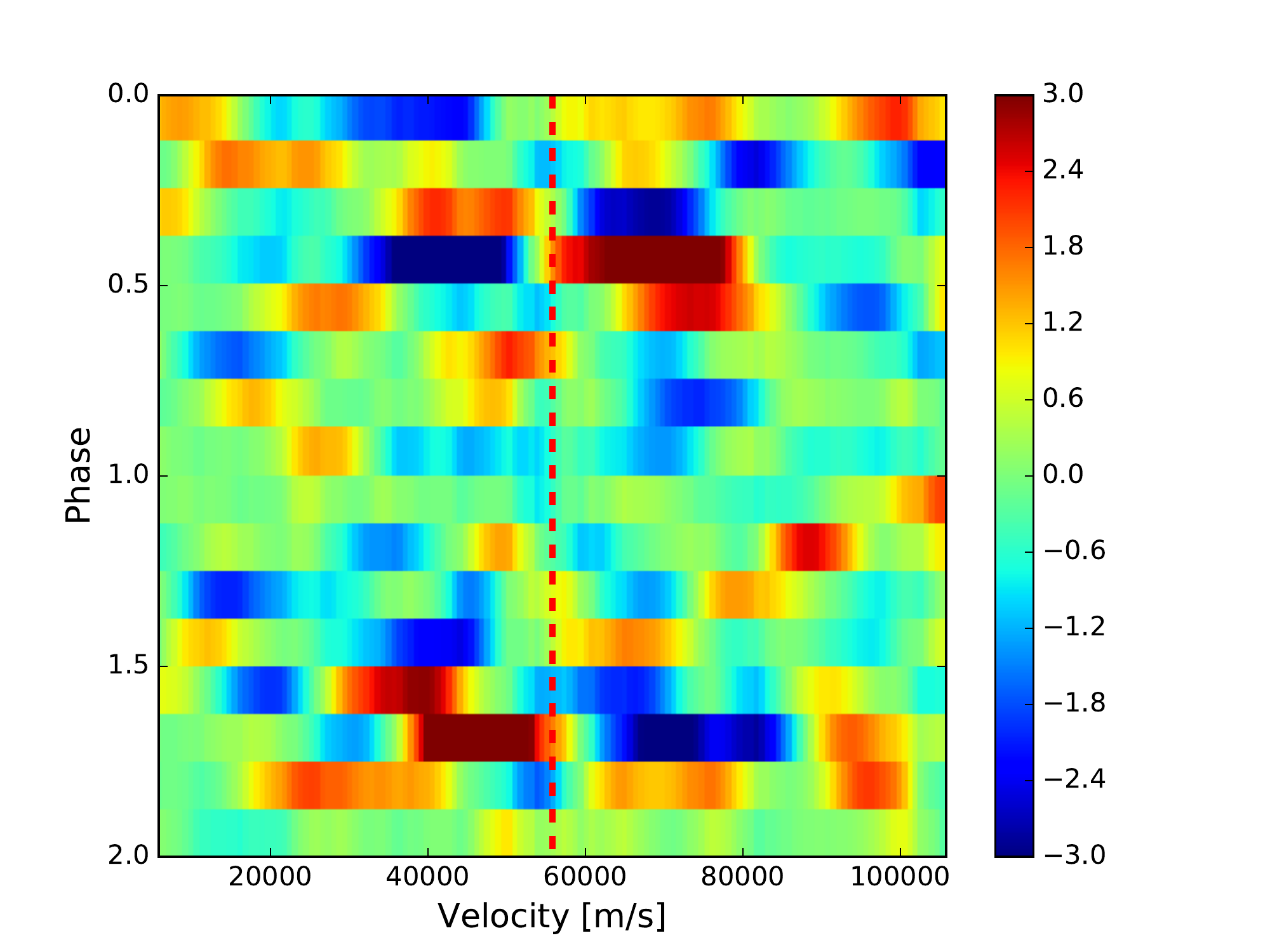} & \includegraphics[width=.4\linewidth]{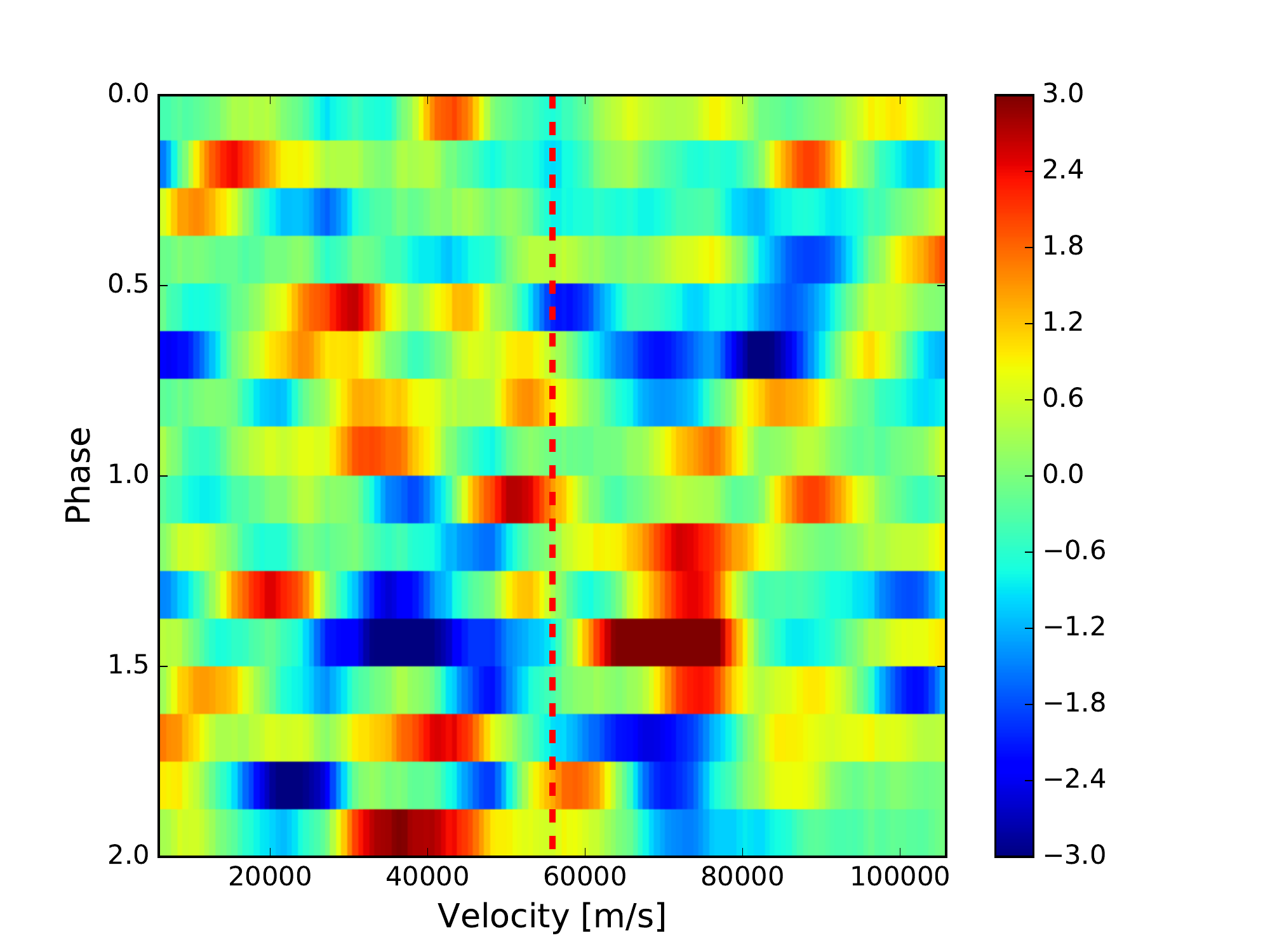} \\
    \includegraphics[width=.4\linewidth]{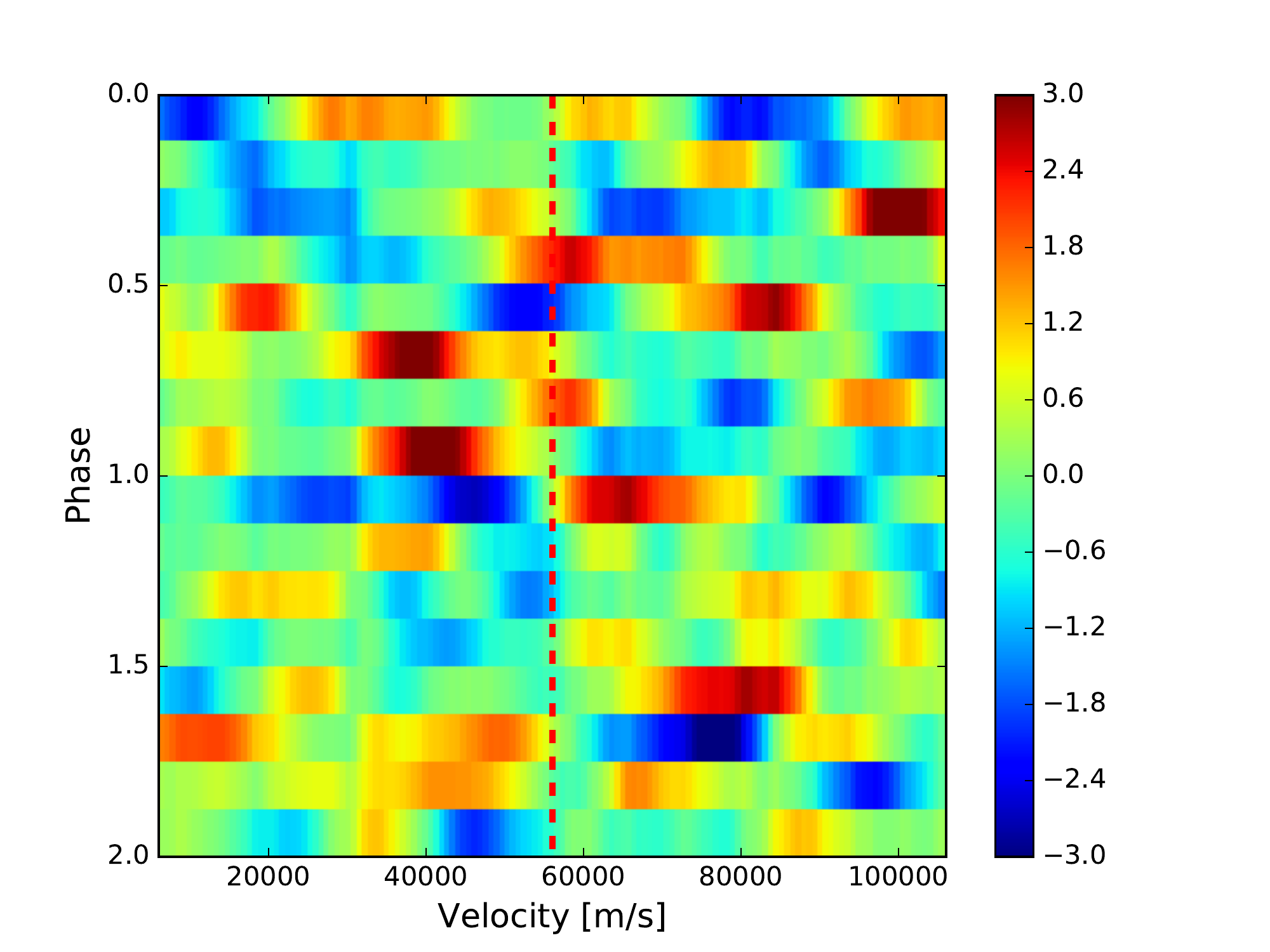} & \includegraphics[width=.4\linewidth]{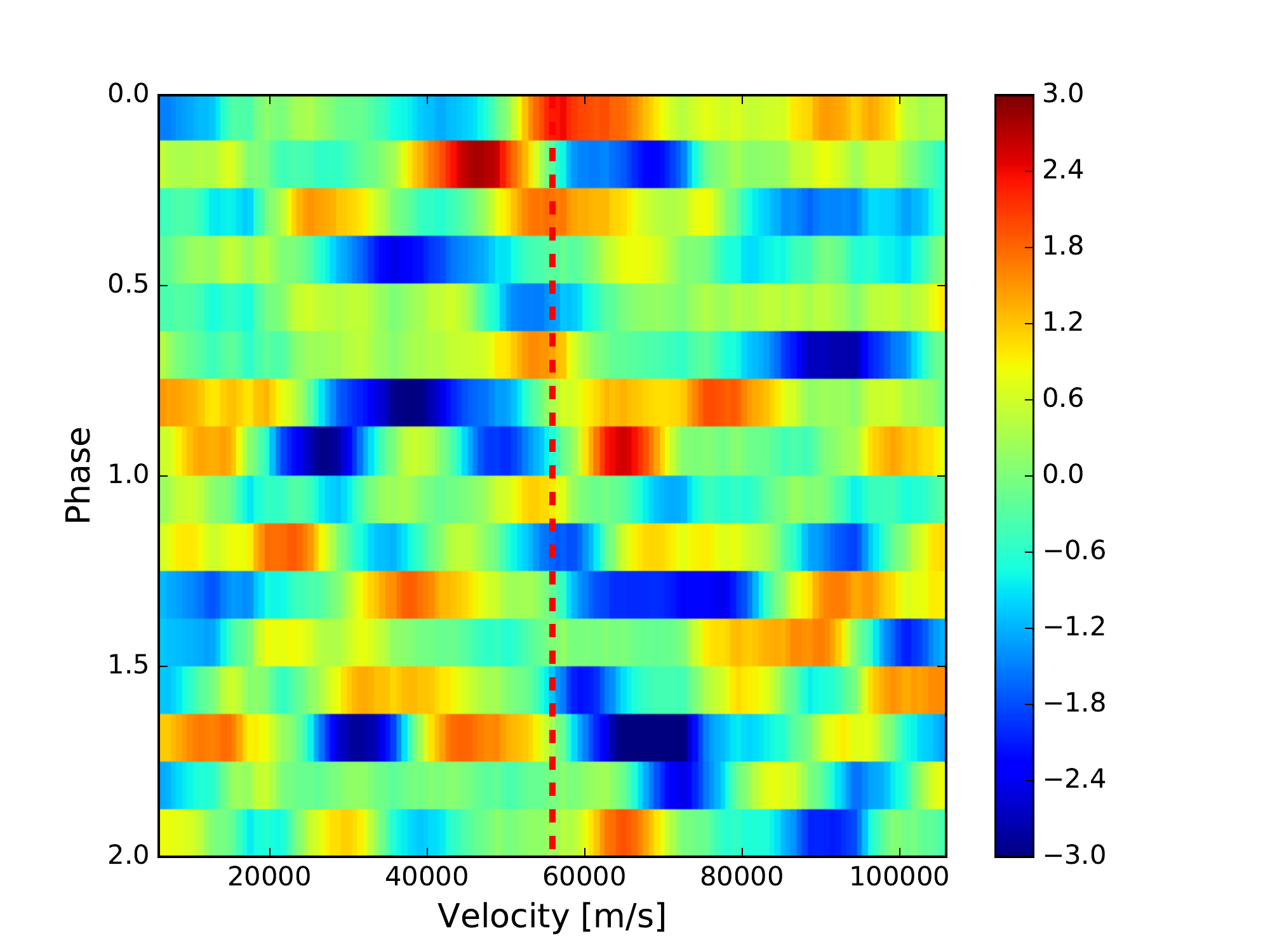} \\
  \end{tabular} \qquad
  \caption{Deviation plots for simulated data for two spot contrasts (0.5 and 0.8). The spot contrast is defined as the relative flux of a spot compared to its surrounding flux. Left (spot contrast = 0.8): the top shows the deviation plot for the noiseless case. From the second panel to bottom, the spot angular radii are $\pi/4$,  $\pi/6$, and $\pi/8$, respectively. Right (spot contrast = 0.5): from top to bottom, the spot angular radii are $\pi/14$, $\pi/16$,  $\pi/18$, and $\pi/20$, respectively. The spot was located on the equator in our simulation. The color scale was set so that red and blue correspond to plus and minus 3-$\sigma$ variations from the median value. The vertical red dashed line indicates the velocity difference between the BD and telluric lines. \label{fig:sim_test_result}}
\end{figure*}

Fig. \ref{fig:sim_test_result} shows the deviation plot as a simulated spot moves across the BD surface. The repeated diagonal blue streak in Fig. \ref{fig:sim_test_result} is the result of the spot sequentially blocking blue-shifted and red-shifted regions. We varied the spot angular radius from $\pi/4$ to $\pi/20$ with a step size of 2 in the denominator. The single spot coverage was thus 25\% and 1.0\% for angular radius of $\pi/4$ and $\pi/20$, respectively. We considered only the single-spot case because the multiple-spot case is a summation process of the single-spot case: the signal of multiple spots would show up in the deviation plot as multiple traces of single spots.

For a spot contrast of 0.8, 6.25\% spot coverage (i.e., $\pi/8$ in angular radius) is difficult to discern with 3-$\sigma$ significance at the noise level we considered (left column in Fig. \ref{fig:sim_test_result}). The significance was calculated by summing up in quadrature the deviation significance along the direction of the spot crossing (see top right panel of Fig. \ref{fig:sim_test_result}). For a spot contrast of 0.5, 1.0\% spot coverage (i.e., $\pi/20$ in angular radius) was difficult to discern with two rotational periods, although the sensitivity to small spot coverage may be increased with multiple-rotation observations. 

%
%

\subsubsection{Tests on Luhman 16 AB data}

\begin{figure}
  \centering
  \begin{tabular}[b]{c}
    \includegraphics[width=.9\linewidth]{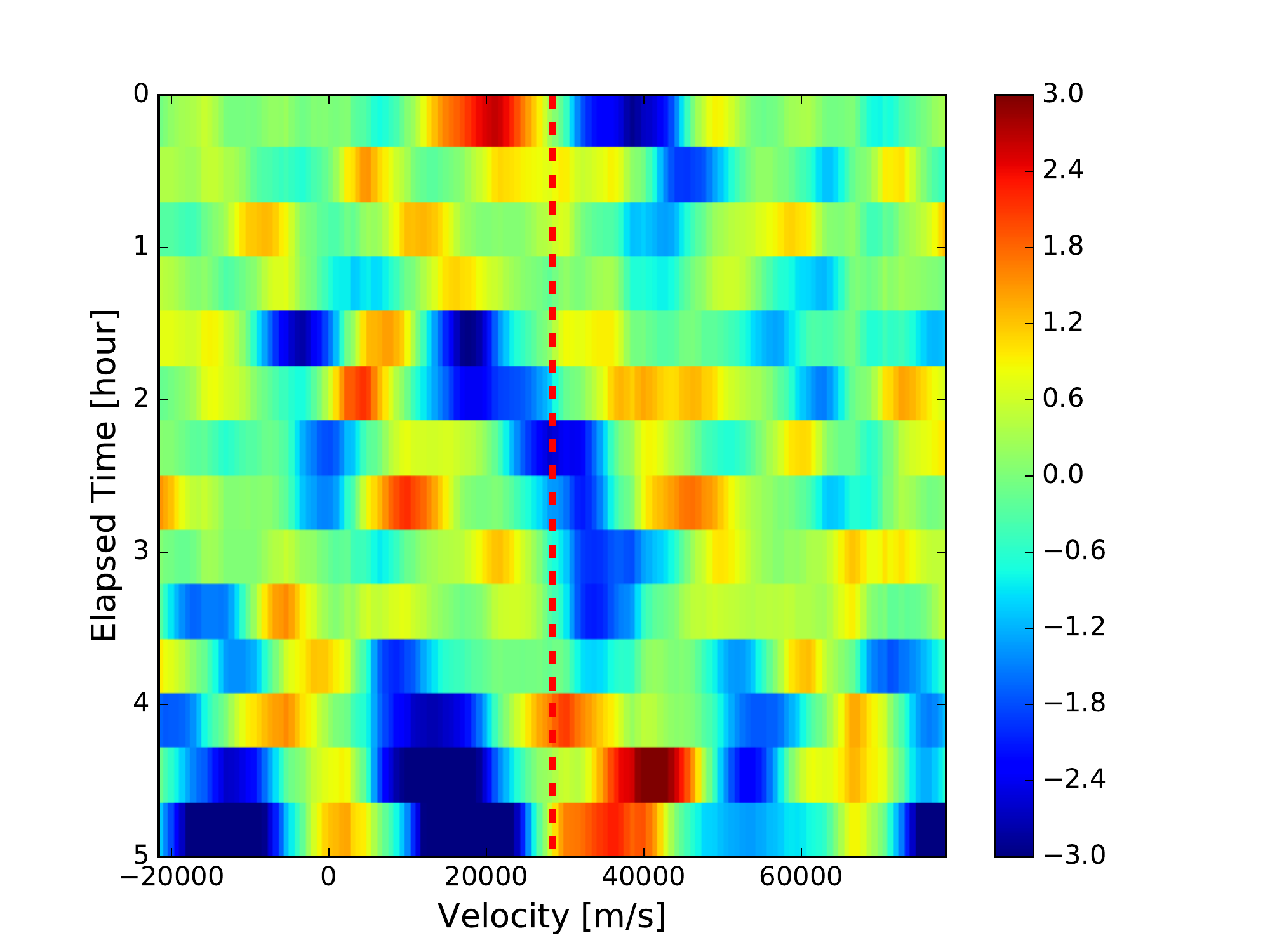} \\
    \includegraphics[width=.9\linewidth]{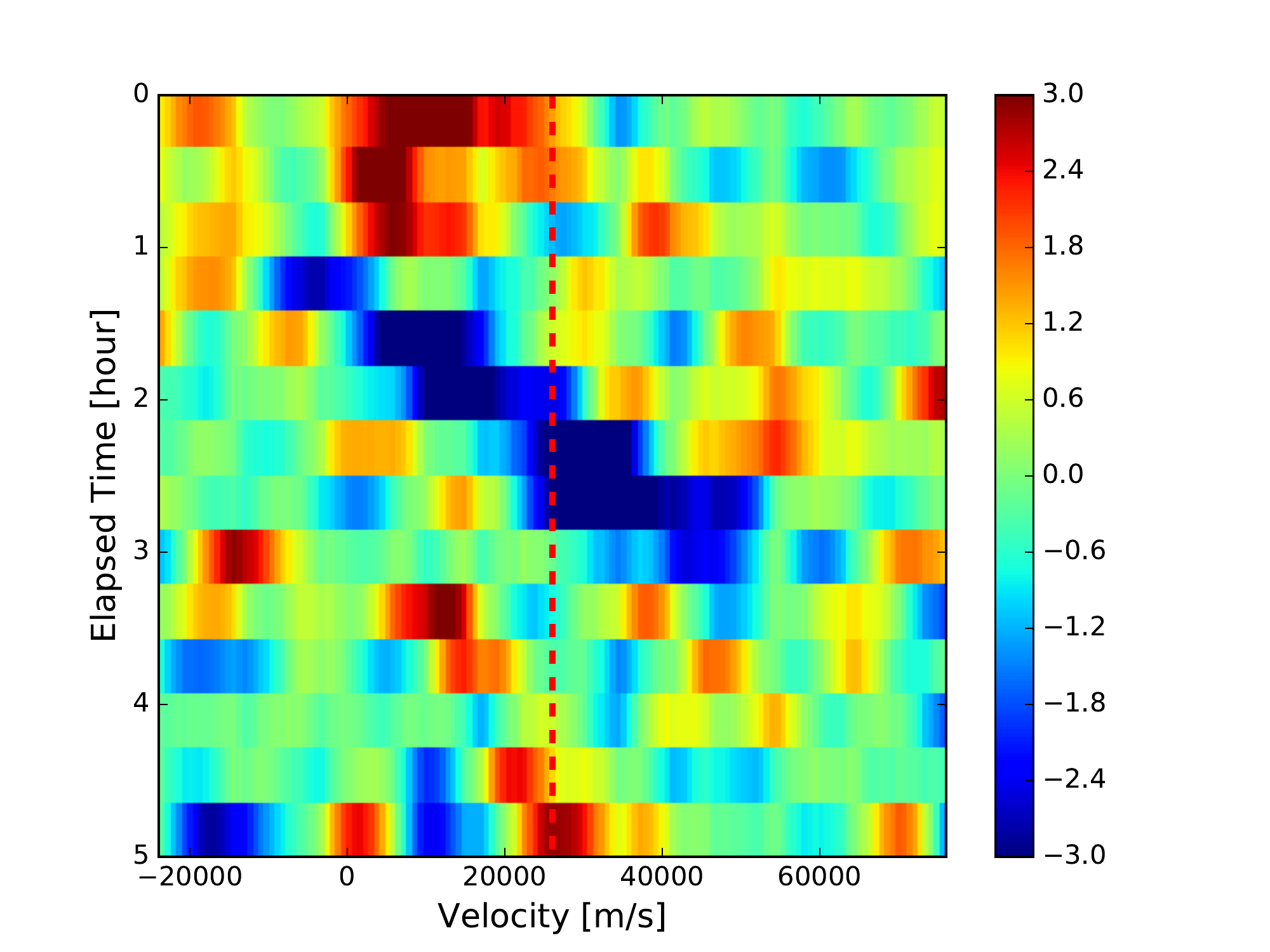} \\
  \end{tabular} \qquad
  \caption{Deviation plots for Luhman 16 A (top) and B (bottom) based on VLT CRIRES data~\citep{Crossfield2014}. The color scale was set so that red and blue correspond to plus and minus 3-$\sigma$ variation from the median value. The vertical red dashed line indicates the velocity difference between the BD and telluric lines.  \label{fig:Luhman_test_result}}
\end{figure}

We tested our data reduction pipeline on published data by analyzing data for Luhman 16 AB from CRIRES on the VLT. We recovered the results demonstrated in~\citet{Crossfield2014}. The line profile deviation plot for Luhman 16 B (Fig. \ref{fig:Luhman_test_result} bottom) shows a strong feature of a spot crossing as reported by~\citet{Crossfield2014}. The line profile deviation plot for Luhman 16 A (Fig. \ref{fig:Luhman_test_result} top) shows no feature more significant than 3-$\sigma$. However, the diagonal feature between 1.5 h and 3.5 h may suggest a spot crossing. 

\begin{figure}
\epsscale{1.1}
\plotone{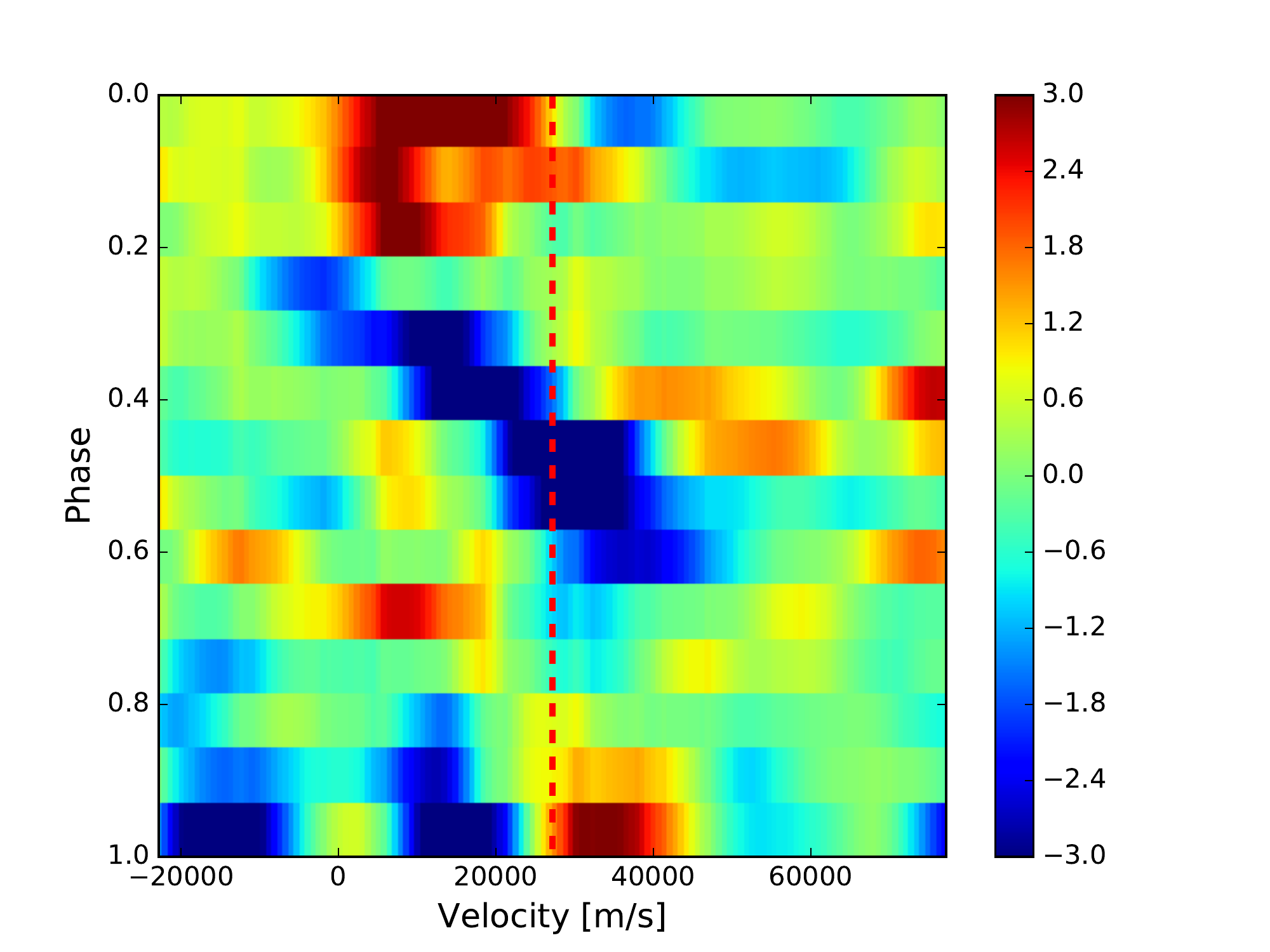}
\caption{Deviation plot for Luhman 16 B based on VLT CRIRES data~\citep{Crossfield2014}. The Luhman 16 B signal is blended with the A signal to mimic seeing-limited observations. The spectral resolution was degraded to R=25,000 to mimic NIRSPEC observations. The color scale was set so that red and blue correspond to plus and minus 3-$\sigma$ variation from the median value. The vertical red dashed line indicates the velocity difference between the BD and telluric lines. \label{fig:Luhman_test_result_mixed}}
\end{figure}

To investigate whether spectral variability of similar magnitude as Luhman 16 B can be detected in seeing-limited Keck NIRSPEC observations, we combined Luhman 16 A and B signal by adding their spectra. Then we degraded the spectral resolution from R=100,000 for VLT CRIRES to R=25,000 for the NIRSPEC data. The spectra were processed by the data analysis package and the result is shown in Fig. \ref{fig:Luhman_test_result_mixed}. With the blended signal from Luhman 16 AB, spectral variability of Luhman 16 B can still be detected without decreased detection significance. The detection of Luhman 16 B in the blended signal is because (1) spectral variability of B is much larger than that of A and (2) A and B are of similar brightness~\citep{Burgasser2014}. 

\subsection{Implications for the J0746 Observations}

J0746 AB have similar brightness in the $K$ band: A is about 40\% brighter than B. Detected spectral variability (if any) should be attributed to the object that dominates the variability (e.g., in the case of Luhman 16 B). However, in the case of a non-detection, we have to rely on simulations to interpret the results. 

Based on our simulations detailed in this section, we can place constraints on the spot size and contrast for J0746. If the spot contrast is 0.5, a single spot coverage of J0746 should be at the $\sim$1\% level or below. Otherwise, the single spot would manifest itself in BD spectral variability. If spot contrast is 0.8, the single spot coverage is $<\sim$6.25\%, which corresponds to a single equatorial spot with an angular radius of $\pi/8$. 

The realistic case is more complicated than the single spot model we adopted in the simulation. There could be multiple spot regions with different sizes and contrasts. However, any spot regions beyond the threshold contrast and size should be detected.

\section{Summary and Discussion}
\label{sec:summary}

We present time-resolved high-spectral resolution observations of J0746 AB, undertaken to search for spectral variability that modulates with rotation. This is the second case study of this type after Luhman 16 AB. We report a null result and place constraints on the surface non-uniformity of J0746 A and B. 

\subsection{Photon Noise vs. Systematical Errors}

The formal uncertainty reported from the data reduction pipeline was 0.5-1.0\% per pixel. After folding the rotation period and combining the spectra for the same phase (see \S \ref{J0746_result}), the uncertainty (standard deviation) per pixel was 1.5-2.0\%, a factor of 2--3 increase compared to the formal uncertainty. This is an indication that the PyNIRSPEC and REDSPEC data reduction pipeline do not produce photon noise limited results. Additional noise may come from the instrumental instability (e.g., pixel shift resulting from temperature, pressure, or mechanical change) and from spectral extraction procedures such as flux interpolation and wavelength calibration. 

Other systematic errors include the imperfect templates used in LSD for retrieving line profiles. The difference between co-phased spectra and reconstructed spectra from line profiles has a RMS of 1.5-3.0\% (see Fig. \ref{fig:lsd_order}), which is higher than the uncertainty of 1.5-2.0\% for the co-phased spectra. In addition, the difference does not resemble Gaussian noise but has structures. The structures are very likely caused by the imperfect templates used for the BD and the Earth's atmosphere. 

For the BD template, any change of effective temperature, surface gravity, metallicity, abundance ratio, and surface non-uniformity will cause the deviation between the template used in LSD and the reality. The amplitude of the deviation is currently unknown. For the telluric template, we use a fixed template generated based on HITRAN data base. In reality, the telluric spectrum changes with time of observation and the location of the object, which is another source of systematics in the data analysis. One could in principle use an evolving telluric spectrum in LSD with a forward modeling approach as in precision RV measurement in NIR. The state-of-the-art for telluric spectrum modeling is at $\sim$0.5-1\% level~\citep{Bean2010, Gao2016}. However, we argue that the major systematics in LSD comes from the imperfect BD template because of all the aforementioned uncertainty in BD atmosphere modeling. 

\subsection{The Prospect of Adaptive Optics Observation}

Another complication in J0746 AB observation is its binary nature. The signal we measured in seeing-limited conditions was the combined flux of A and B. Phase folding observed spectra based on rotational period will strengthen the signal of one and weaken the signal of the other. However, given the limited duration of observation, we had slightly more than two periods for A and more than three periods for B. There may be still contaminating signal from one to the other. A better way to observe J0746 AB is to use NIRSPEC in adaptive optics (AO) mode. In this mode, the system will be spatially resolved before being dispersed by a spectrograph. The angular separation between A and B is 0.12$^{\prime\prime}$-0.36$^{\prime\prime}$ at periastron and apastron, which can be resolved by the Keck telescope in $K$ band ($\lambda$/D=40 mas). Flux contamination will be much alleviated in Keck AO observations. Therefore, future AO-aided high spectral resolution observations will be helpful for observing tight binary BDs like J0746 AB.  



\noindent{\it Acknowledgements} 
We would like to thank Ian Crossfield for sharing the VLT CRIRES data for Luhman 16 AB and for constructive discussions on BD spectral variability and Doppoler imaging. We would like to thank Courtney Dressing for her suggestion of binning CRIRES data to investigate the impact of reduced spectral resolution on Keck NIRSPEC data.

\bibliography{mybib_JW_DF_PH5}

\clearpage

\input{J0746.tex} 
\end{document}

%% file: J0746.tex

\begin{deluxetable}{lllll}
\tablewidth{0pt}
\tablecaption{Properties of 2MASSW J0746425+200032AB.\label{tab:J0746}}
\tablehead{
\colhead{\textbf{Parameter}} &
\colhead{\textbf{Unit}} &
\colhead{\textbf{A}} &
\colhead{\textbf{B}} &
\colhead{\textbf{Ref.}} \\
}

\startdata
Distance & pc & \multicolumn{2}{c}{12.21$\pm$0.05}  & ~\citet{Dahn2002} \\
Orbital Period & yrs & \multicolumn{2}{c}{12.71$\pm$0.07}  & ~\citet{Konopacky2010} \\
Semi-major Axis & mas & \multicolumn{2}{c}{$237^{+1.5}_{-0.4}$}  & ~\citet{Konopacky2010} \\
Total Mass & $M_\odot$ & \multicolumn{2}{c}{0.151$\pm$0.003}  & ~\citet{Konopacky2010} \\
Eccentricity & \nodata & \multicolumn{2}{c}{0.487$\pm$0.003}  & ~\citet{Konopacky2010} \\
$T_0$ & yrs & \multicolumn{2}{c}{2002.83$\pm$0.01}  & ~\citet{Konopacky2010} \\
Orbital Inclination & degrees & \multicolumn{2}{c}{138.2$\pm$0.5}  & ~\citet{Konopacky2010} \\
System RV & $\rm{km}\ \rm{s}^{-1}$ & \multicolumn{2}{c}{54.7$\pm$0.8} & ~\citet{Konopacky2010} \\
Mass Ratio & \nodata & \multicolumn{2}{c}{$4.0^{+0.1}_{-3.8}$} & ~\citet{Konopacky2010} \\

$M_J$ & mag & 11.85$\pm$0.04 & 12.36$\pm$0.10 & ~\citet{Konopacky2010} \\
$M_H$ & mag & 11.13$\pm$0.02 & 11.57$\pm$0.03 & ~\citet{Konopacky2010} \\
$M_{K_P}$ & mag & 10.62$\pm$0.02 & 10.98$\pm$0.02 & ~\citet{Konopacky2010} \\
Mass & $M_\odot$ & $0.12^{+0.01}_{-0.09}$ & $0.03^{+0.09}_{-0.01}$ & ~\citet{Konopacky2010}  \\
Spectra Type & \nodata & L0$\pm$0.5 & L1.5$\pm$0.5 & ~\citet{Bouy2004}  \\
T$_{\rm{eff}}$ & K & 2205$\pm$50 & 2060$\pm$70 & ~\citet{Konopacky2010} \\
Radius & R$_{\rm{Jupiter}}$ & 0.99$\pm$0.03 & 0.97$\pm$0.06 & ~\citet{Konopacky2010} \\
Rotation Period & hour & 3.32$\pm$0.15 & 2.07$\pm$0.002 & ~\citet{Harding2013},~\citet{Berger2009}  \\
V$\sin i$ & $\rm{km}\ \rm{s}^{-1}$ & 19$\pm$2 & 33$\pm$3 & ~\citet{Konopacky2012} \\
Equatorial Velocity & $\rm{km}\ \rm{s}^{-1}$ & 36$\pm$4 & 56$\pm$2 & ~\citet{Harding2013} \\
Equatorial Inclination$^{\ast}$ & degrees & 32$\pm$4 & 36$\pm$4 & ~\citet{Harding2013} \\

\enddata

\tablecomments{$\ast$: can be 90 degrees off because of the uncertainty of a pro-grade or a retrograde orbit. }

\end{deluxetable}